\documentclass[12pt]{article}
\pdfoutput=1
\usepackage{jheppub}
\usepackage[utf8]{inputenc}
\usepackage{verbatim}
\usepackage{amsmath}
\usepackage{amssymb}
\usepackage{amsthm}
\usepackage{amsfonts}
\usepackage{hyperref}
\usepackage{physics}
\usepackage{float}
\usepackage{xcolor}
\usepackage{graphicx}
\usepackage{simpler-wick}

\newcommand{\be}{\begin{equation}}
\newcommand{\ee}{\end{equation}}
\newcommand{\ba}{\begin{align}}
\newcommand{\ea}{\end{align}}

\newcommand{\ol}{\overline}

\newcommand{\bi}{\begin{itemize}}
\newcommand{\ei}{\end{itemize}}
\newcommand{\bfig}{\begin{figure}\begin{center}}
\newcommand{\efig}{\end{center}\end{figure}}

\newtheorem{thm}{Theorem}

\newcommand{\cA}{\mathcal{A}}
\newcommand{\cH}{\mathcal{H}}

\def\ie{i.e.~}
\def\ket#1{| #1 \rangle}
\def\bra#1{\langle  #1 |}

\def\TFD{{\mathrm{TFD}}}
\def\hat{\widehat}
\def\b{{\sf b}}
\def\a{{\sf a}}
\def\t{\widetilde }
\def\h{\widehat}
\def\spc{\hspace{1pt}}

\newcommand{\rmd}{\mathrm d}

\begin{document}
\title{The mini-Page Curve in Cosmology}
\author[\phi]{Ricardo Esp\'indola}
\author[\bar{\phi}]{and Shoichiro Miyashita}
\affiliation[{\phi}]{Institute for Advanced Study, Tsinghua University, Beijing 100084, China}
\affiliation[\bar{\phi}]{Department of Liberal Arts and Sciences,
Faculty of Engineering, Takushoku University, 
Hachioji, Tokyo 193-0985, Japan}
\emailAdd{ricardo.esro1@gmail.com,\ miyashita1miyashita@gmail.com}
\abstract{The black hole information paradox has motivated extensive study of how and when information escapes from evaporating black holes. Here we address the analogous question for cosmological horizons: when does an individual Hawking pair begin to carry information out of a de Sitter horizon? We study this problem in a class of two-dimensional flow geometries that interpolate smoothly between an asymptotic AdS$ _2 $ boundary and a dS$ _2 $ static patch. Modeling the emission of a Hawking pair via a probe state constructed from local operators and their modular conjugates, we promote the centaur algebra of observables to a Type II$ _\infty $ factor through the crossed-product construction. This allows us to compute the entropy difference between the thermofield-double reference state and the Hawking-pair state. We find that this difference traces a characteristic mini-Page curve for the cosmological horizon: it starts near zero, reaches a minimum near $ \tau \approx \beta/8 $ before increasing again. We interpret the location of the minimum as the time at which quantum information begins to escape the cosmological horizon. Extending the analysis to the microcanonical ensemble, we show that the algebraic entropy coincides with the generalized entropy of an entanglement wedge cut that tracks the emitted particle along the horizon. Furthermore, the relative modular flow generated between the two states yields a Lyapunov exponent $\lambda =2\pi/\beta$, identifying the scrambling time as the scale at which the information carried by the pair becomes accessible to a static-patch observer.
}

\maketitle

\section{Introduction}

The black hole information paradox \cite{Hawking:1976ra} remains as one of the sharpest diagnostics of the tension between quantum mechanics and gravity. Hawking radiation \cite{Hawking:1975vcx} emitted by evaporating black holes appears to be thermal for an asymptotic observer, carrying no information about the black hole interior. This produces a monotonic increase in the von Neumann entropy of the exterior radiation that eventually exceeds
\be
S_{\rm BH} = \frac{A}{4G}~,
\ee
the Bekenstein--Hawking entropy \cite{Bekenstein:1972tm, Bekenstein:1973ur, Hawking:1975vcx}. 
Unitarity of the underlying quantum field theory, however,  demands that the radiation purify at late times. As a consequence of novel wormhole contributions to  the gravitational path integral, the entropy of Hawking radiation follows the celebrated Page curve \cite{Page:1993df, Page:1993up, Page:1993wv} according to the unitarity expectation: the entropy rises linearly until the Page time, after which an island \cite{Penington:2019npb, Almheiri:2019psf, Almheiri:2019hni, Almheiri:2019yqk, Hartman:2020khs, Espindola:2022fqb, Bousso:2022gth, Bousso:2023kdj, Antonini:2025sur} emerges and the entropy decreases, returning to zero once the black hole has evaporated \cite{Almheiri:2019qdq, Penington:2019kki}. 

While enormous progress has been made on recovering the Page curve from the gravitational semiclassical action for evaporating black holes and on understanding how information from the black hole interior is encoded in the radiation system \cite{Brown:2019rox, Chen:2019iro,Espindola:2025ons}--notably through the discovery of entanglement islands and replica wormholes, which imply that the island is the hologram of radiation--the question of when this transfer occurs has received comparatively less attention. In particular, a precise microscopic characterization of the timescale at which an individual Hawking pair begins to carry quantum information out of the black hole interior remains elusive.  

Recently, a powerful algebraic probe designed for this purpose has been developed \cite{Verlinde:2022xkw}. Within this framework the thermofield-double (TFD) state serves as a reference, while a probe ``Hawking-pair state'' is constructed by acting with a local operator together with its partner. The partner is modeled using the modular conjugate mirror operator. The entropy difference between these states isolates the quantum information carried by the pair. Computations in this setting have already led to mini-Page curves for asymptotically AdS black holes, revealing a characteristic timescale set by the inverse temperature at which information begins to leak.

Cosmological horizons present a sharper challenge. Gibbons and Hawking showed that cosmological horizons obey a similar, yet even more profound, area-entropy relation \cite{Gibbons:1977mu}. A static patch observer in de Sitter (dS) space detects thermal radiation emanating from the cosmological horizon. Unlike black hole horizons, the entropy of a cosmological horizon is intrinsically tied to the observer who measures it, leading to a radical departure from the black hole case. Distinct observers in cosmology perceive different horizons. A controlled arena in which Hawking radiation from cosmological horizons can be analyzed is provided by flow geometries \cite{Anninos:2017hhn, Anninos:2018svg}. These are two-dimensional dilaton-gravity theories that interpolate smoothly between an asymptotic AdS$_2$ boundary and a dS$_2$ static patch in the deep bulk interior. Among these solutions, the sharp centaur geometry provides an exact background solution in which the AdS and dS regions are glued at a smooth region, permitting a precise definition of bulk operators and correlation functions that probe the interior dS space using standard tools of AdS/CFT \cite{tHooft:1993dmi, Susskind:1994vu, Maldacena:1997re, Witten:1998qj, Gubser:1998bc}. 

\begin{figure}[t]
\begin{center}
\includegraphics[width=7.cm]{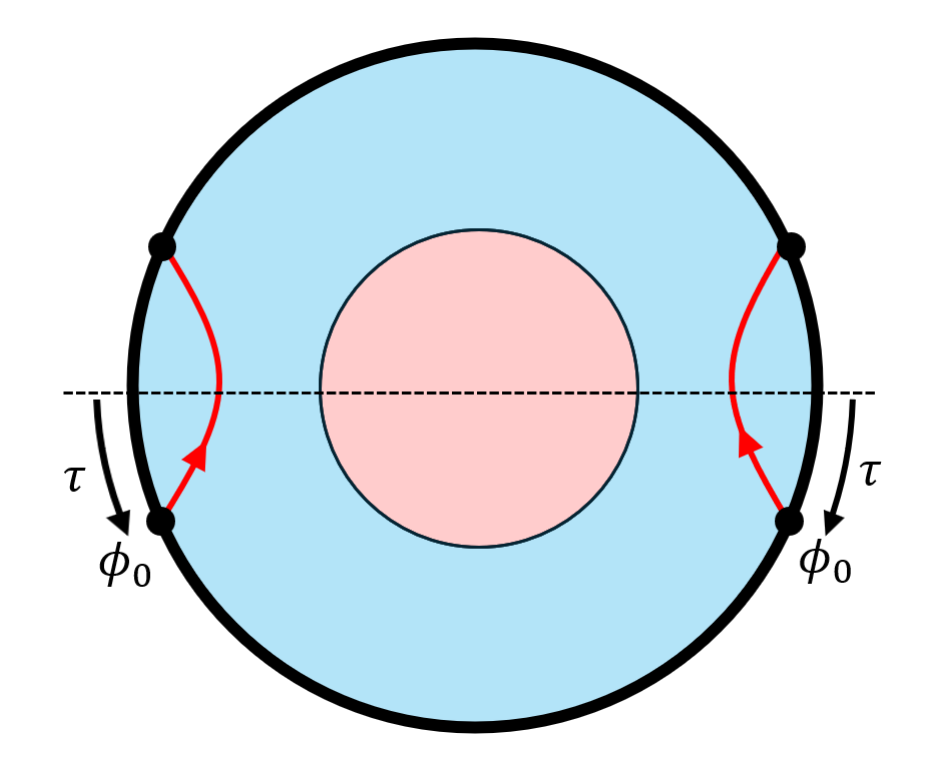} 
	\caption{Euclidean centaur geometry with a Hawking pair. One member of the pair is created by an operator $\phi_{0}$ acting on the past (and future) ``right" boundary, while the other is created by the same operator acting on the past (and future) ``left" boundary. In the Lorentzian picture, the former is interpreted as a local bulk operator and the latter as its mirror operator. Their positions are given by the intersections of the Euclidean trajectories with the time-symmetric slice.
    }
\label{fig:ECWP}
\end{center}
\end{figure}

In this work we analyze the question of when information is transmitted from the dS horizon. To this end we study the Hawking pair model in the flow geometry framework. The state associated  with a Hawking pair is modeled as a conjugate pair of particles propagating in Euclidean space, which includes both the AdS and dS regions, and is constructed explicitly via modular conjugation (see Fig. \ref{fig:ECWP}). 
We employ the tools of algebraic quantum field theory, in particular the centaur-algebra of observables together with its mirror operators \cite{Aguilar-Gutierrez:2023odp}. This framework allows us to define and compute the entropy change induced by the presence of the Hawking particles by promoting the centaur-algebra to a Type II$_\infty$ factor via the crossed product construction \cite{Witten:2021unn} and by using the holographic dictionary for the centaur geometry.

We present a summary of our main results. In Sec. \ref{sec:basic} we review the basic framework. We begin with the class of two-dimensional flow geometries that interpolate between an asymptotic AdS$_2$ boundary and a dS$_2$ static patch in the infrared. We specialize to the sharp centaur solution, which is used as the gravitational background for the subsequent analysis. We then recall the essential elements of Tomita–Takesaki modular theory \cite{Takesaki:1970aki} and von Neumann algebras that will be used throughout the paper. 

In Sec. \ref{sec:3} we develop the algebraic framework required for our analysis. We introduce the centaur-algebra of observables associated with the right exterior of the flow geometry, a Type III$_1$ von Neumann algebra generated by matter fluctuations. We then construct the Hawking-pair state by acting on the thermofield-double reference state with a local operator together with its mirror operator
\be
\ket \Phi := N^{-1/2} \phi_\tau \bar{\phi}_\tau \ket \Psi~.
\ee
Finally, we promote the algebra to a Type II$  _\infty$ factor via the crossed-product construction. This step provides a well-defined trace, density matrix and allows us to cleanly separate the gravitational and field-theoretic contributions to the entropy difference $\Delta S$ between the Hawking and background states.

We then derive Euclidean boundary two-point functions in the centaur geometry in Sec. \ref{sec:twopoint}. In the large mass limit, we study geodesics in the AdS region and those that enter the dS-hemisphere. We show that the dominant contributions to the correlators come from trajectories that remain entirely in the AdS region. The resulting two-point function given by
\be
\label{centaur2pt}
G_{\beta}(\Delta \tau) \simeq \left( \frac{2\pi \ell^2 \widetilde{\Phi}_{b} }{\mathcal{C} \beta} \frac{1}{\sinh( \frac{\pi \Delta \tau}{\beta})} \right)^{2 m \ell} ~,
\ee
which incorporates the effects of the dS interior through the geometry, serves as the key input for the entropy computations that follow in the next section.

We next evaluate the entropy difference $\Delta S$ between the thermofield-double reference state and the Hawking-pair state in detail in Sec. \ref{sec:5}. Working within the crossed-product Type II$_\infty$ algebra, we compute the gravitational contribution using the replica trick and extract the field-theoretic contribution via Araki’s formula. It is important to emphasize that $ \Delta S_{\rm grav}$ (and thus $\Delta S$) captures only the contribution of this single probe state. The entropy of the global cosmological horizon is determined by the total energy and matter content in the static patch through metric backreaction; one Hawking pair produces only a small, localized effect whose physics we are interested in. 

The resulting expression for $\Delta S$ as a function of the Euclidean-time separation of the Hawking pairs $\Delta\tau$ exhibits the following behavior. The curve starts near zero, decreases, and reaches a minimum around the midpoint $\Delta\tau \approx \beta/8$. Beyond this point $\Delta S$ increases again and approaches zero. This behavior is opposite to the standard black-hole Page curve and constitutes what we dubbed an \emph{inverse mini-Page curve} for the cosmological horizon. The inverse character originates from the distinctive thermodynamics of the cosmological horizon in the centaur geometry. The creation and separation of the Hawking pair acts as a matter fluctuation that backreacts on the cosmological horizon. Because of the effective negative specific heat associated with the dS region, this backreaction initially reduces the gravitational entropy contribution. After the minimum, further separation no longer produces significant additional backreaction on the horizon, so the entropy contribution saturates and begins to increase. The location of the minimum marks the precise microscopic timescale at which quantum information begins to be carried out of the dS horizon by the Hawking pair.

We conclude our technical analysis in Sec. \ref{sec:Sgen} by extending the framework to the microcanonical ensemble and computing the generalized entropy. We work with the microcanonical TFD state and the associated crossed-product Type II$_\infty$ algebra. We study the Hawking-pair labeled by the outgoing null coordinate on the future cosmological horizon. We show that the resulting algebraic von Neumann entropy on this algebra is precisely equal to the generalized entropy of the entanglement wedge cut that tracks the emitted particle along the horizon. We then examine the real-time dynamics generated by the relative modular operator between the Hawking-pair state and the reference state. This produces an effective relative modular flow that acts on horizon operators as a null-coordinate shift. The exponential growth of this shift produces a Lyapunov exponent of $\lambda = 2\pi/\beta$ and identifies the scrambling time as the timescale after which the quantum information carried by the pair becomes accessible to a static-patch observer. Together, these results provide an algebraic realization of information transfer across the cosmological horizon in flow geometries.

Finally, several calculations are collected in the appendices. In Appendix \ref{App:JTtwopt} we review the derivation of the Euclidean boundary two-point function in pure JT gravity. This serves both as a consistency check and as a reference point for the more involved geodesic analysis performed in the centaur geometry. Appendix \ref{App:crossed} provides a self-contained review of the crossed-product construction. We discuss its group-theoretic foundations and its algebraic implementation, which are used to promote the centaur-algebra to a Type II$  _\infty  $ factor and to define the trace and entropy functionals employed in the main text.

\section{Basic framework}
\label{sec:basic}
In this section, we review the basics and set notation for the rest of the paper. In Sec. \ref{sec:flow} we introduce the class of two-dimensional flow geometries that interpolate from an asymptotic AdS$_2$ boundary to a dS$_2$ static patch in the deep interior. In particular the sharp centaur solution provides the gravitational background for all subsequent analysis. In Sec. \ref{sec:modular} we develop the necessary operator-algebraic framework. We first recall the general concepts of von Neumann algebras and states in Sec. \ref{sec:toolkit}, and then present in Sec. \ref{sec:Tomita} the key elements of Tomita–Takesaki modular theory. For a more complete treatment of the subject, we refer to \cite{Witten:2018zxz, Liu:2025krl}.  

\subsection{Flow geometries}
\label{sec:flow}
We consider the two-dimensional dilaton-gravity theory whose Euclidean dynamics is governed by the action 
\begin{equation}\label{eq:I centaur}
    I_E=-\frac{1}{16\pi G_2}\int_{\mathcal{M}} \rmd^2x\sqrt{g}(R\Phi+V(\Phi))-\frac{1}{8\pi G_2}\int_{\partial\mathcal{M}} \rmd \tau\sqrt{h}\,\Phi K+I_{\text{ct}}~,
\end{equation}
where $\cal{M}$ denotes the spacetime manifold, $h$ is the induced metric on the boundary $\partial \cal{M}$, and the counterterm takes the explicit form
\be 
\label{eq:IctJT}
I_{\text{ct}}
=\frac{1}{8\pi G_{2}}\int_{\partial\mathcal{M}} \rmd \tau\sqrt{h}\frac{\Phi}{\ell}~.
\ee
The parameter $\ell$ is fixed by the dilaton potential $V(\Phi)$. It is also possible to include in the action a topological term
\begin{align}
    I_{\text{top}}&=-\frac{\Phi_{0}}{16\pi G_{2}}\int_{\mathcal{M}}\rmd^{2}x\sqrt{g}R-\frac{\Phi_{0}}{8\pi G_{2}}\int_{\partial\mathcal{M}}\rmd\tau\sqrt{h}K=-\frac{2\pi}{8\pi G_{2}}\Phi_{0}\chi\;,\label{eq:topterm}
\end{align}
with $\chi$ the Euler characteristic of $\mathcal{M}$.

\begin{figure}[t]
\begin{center}
\includegraphics[width=13.cm]{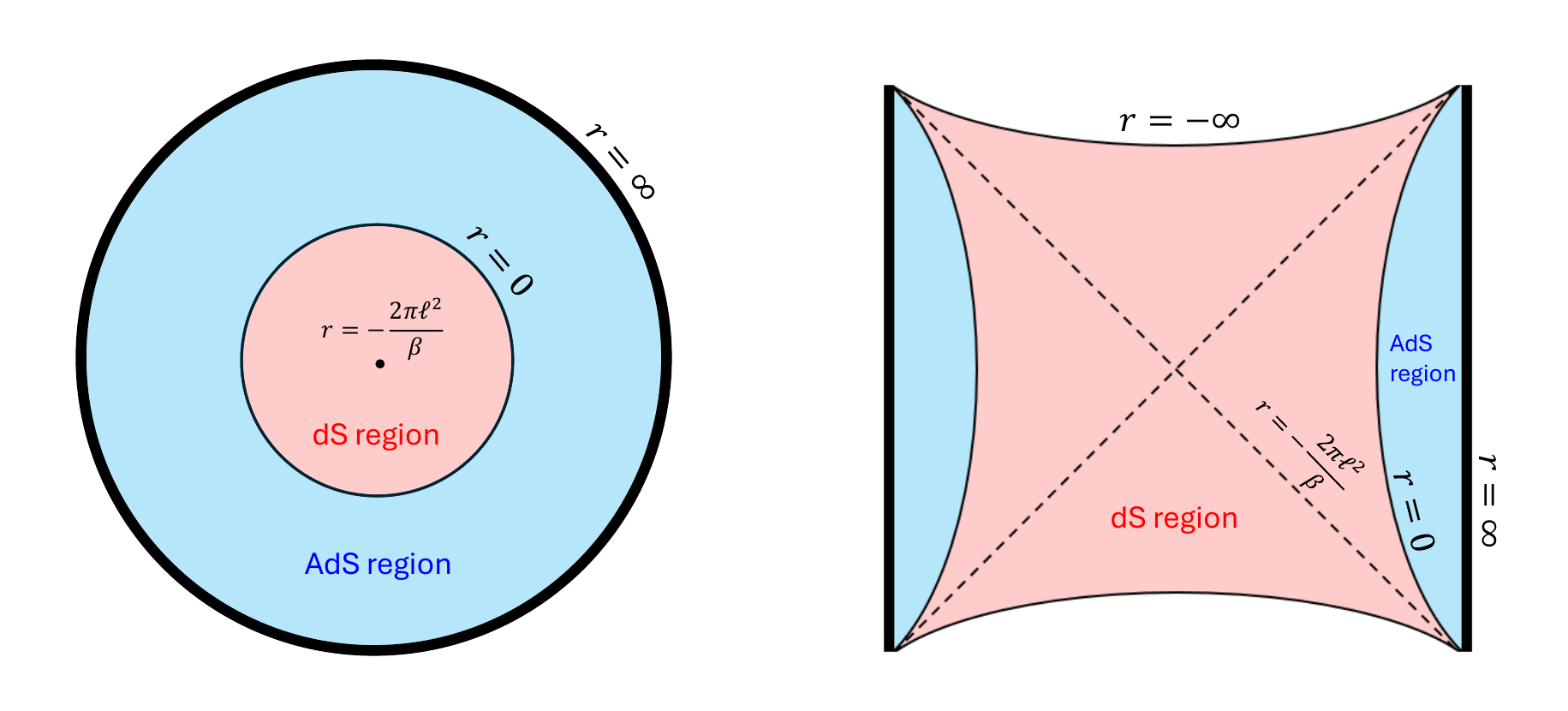} 
	\caption{(Left) Euclidean sharp centaur geometry \eqref{eq:sharpE}.  (Right) Lorentzian sharp centaur geometry. 
    }
\label{fig:centaur1}
\end{center}
\end{figure}

The dilaton field and the metric satisfy the coupled equations of motion
\be
\label{eq:EOM}
\nabla_\mu\nabla_\nu\Phi-g_{\mu\nu}\nabla^2\Phi-\frac{1}{2}g_{\mu\nu}V(\Phi)=0~, \quad R=-V'(\Phi)~.
\ee
The gravity equation controls the dilaton profile, which in turn selects the background geometry $\mathcal{M}$. A convenient family of static solutions can be written in Schwarzschild gauge as
\begin{equation}
\label{eq:static patch sol}
    \rmd s^2=N(r)\rmd\tau^2+\frac{\rmd r^2}{N(r)}~,\quad \Phi=\widetilde{\Phi}_{b} \frac{r}{\ell}~,
\end{equation}
where $\widetilde{\Phi}_{b}>0$ is a dimensionless normalization constant. Throughout this work we impose $\Phi_{0}\gg \widetilde{\Phi}_{b}$; from the higher dimensional perspective, this condition corresponds to a near-extremal configuration. In this coordinate system the Ricci scalar is position-dependent, $R=-\partial^{2}_{r}N(r)$, and
\begin{equation}\label{eq:N(r) factor}
    N(r)=\frac{\ell}{\widetilde{\Phi}_{b}}\int_{r_h}^r d r'\,V(r')~. 
\end{equation}
The horizon sits at the coordinate $r = r_h$, where the function $N(r)$ vanishes. Smoothness of the Euclidean geometry near this point requires that the time coordinate $\tau$ be identified periodically with inverse temperature
\be
\beta:= \frac{1}{T} = \frac{4\pi \widetilde{\Phi}_{b}}{\ell|V(r_{h})|}~, \label{eq:flowtemp}
\ee 
For the concrete calculations in this paper we specialize to the \emph{sharp centaur} geometry \cite{Anninos:2017hhn, Anninos:2018svg} (Fig. \ref{fig:centaur1})
\be
\label{eq:sharpE}
ds^2 =  \left( \frac{4\pi^2 \ell^2}{\beta^2}+ \frac{1}{\ell^2}\frac{r^3}{\lvert r \lvert} \right) d\tau^2 + \frac{dr^2}{\left( \frac{4\pi^2 \ell^2}{\beta^2}+\frac{1}{\ell^2}\frac{r^3}{\lvert r \lvert} \right)}~ , \quad - \frac{2\pi \ell^2}{\beta} \leq r < \infty~. 
\ee
This background is an exact solution when the dilaton potential is chosen as $V(\Phi)=2|\Phi|/\ell^2$ 
and horizon radius is at $r_{h} = -2 \pi \ell^2 /\beta$. The metric is continuous and twice-differentiable everywhere. Evaluating the dilaton on the horizon yields $\Phi_{h} = \Phi(r_{h}) = -2\pi \ell  \widetilde{\Phi}_{b}/\beta$, so that the absolute value $|\Phi_{h}|$ directly sets the temperature of the system. For any given temperature $T$ the theory admits  two class of solutions: one with negative $\Phi_h$ (corresponding to a dS$_2$ cosmological horizon) and the one with positive $\Phi_h$ (corresponding to an AdS$_2$ black hole).

The sharp centaur arises as the limiting case of a larger family of  solutions generated by smooth potentials that are parametrized by a single transition scale $\epsilon$. A representative example is $V(\Phi) = (2/\ell^2) \Phi \tanh(\Phi/\epsilon)$. The boundary theory is controlled by the Schwarzian action, whose variation gives the Hamiltonian
\be
H = \frac{\phi_r(u)}{8 \pi G_N} \left( \frac{t'(u)^2}{2} - \frac{t'''(u)}{t'(u)} + \frac{3}{2} \left( \frac{t''(u)}{t'(u)}\right)^2 \right)~,
\ee
where $t(u)$ denotes the reparameterization of boundary proper time.

\subsection{Modular theory}
\label{sec:modular}
The natural mathematical framework for analyzing systems with an infinite number of degrees of freedom is Tomita-Takesaki theory. Before turning to the details of that theory, we first review a few general concepts from operator algebras, in particular von Neumann algebras, which will be useful throughout the rest of the paper. 

\subsubsection{Algebraic toolkit}
\label{sec:toolkit}
The operator-algebra framework has an extremely wide range of applicability. Its power comes from a surprisingly small set of deep unifying concepts. Let $\widetilde\cH$ be a Hilbert space, possibly non-separable (\ie, of uncountable dimension). We can define a separable Hilbert space $\cH$ contained in $\widetilde{\cH}$ by restricting our attention to a local algebra of operators. In order to specify $\cH$  we need two ingredients: the class of observables and the class of states we are interested in. Intuitively, we motivate such an algebra by restricting to a set of operators relevant in a certain spacetime region. More precisely, the set of operators $\a$ to which we have access forms an algebra that we denote by $\cA_0$. Apart from the basic properties of an algebra, we work throughout with unital algebras, meaning that there exists an element ${\bf 1}\in \cA$ such that ${\bf 1}~\a = \a ~ {\bf 1} = \a$, $\forall~\a \in \cA$. In contrast to the standard approach to quantum systems, where knowledge of the global state is required, this framework works with a particular state of interest together with the set of operators that act on it, therefore trading global information for local information. We denote the state by $\ket{\psi}$.

By acting with operators $\a \in\cA$ on the state we obtain the so-called pre-Hilbert space $\cH_0 = \{~ \a \ket{\psi}~, \forall~ \a \in \cA_0~ \}$. In general this space can be infinite dimensional, but in comparison with $\widetilde\cH$ it is separable. We will be interested in $*$-algebras, meaning that for any operator $\a \in \cA_0$, there is a corresponding $\a^* \in \cA_0$. In physics we denote the $*$-operation by the dagger. The previous procedure defines a pre-Hilbert space $\cH_0$, \ie, a space that satisfies all the axioms of a Hilbert space except completeness. To obtain a genuine Hilbert space, we complete $\cH_0$ by including all Cauchy sequences in $\cH_0$. We also complete the algebra $\cA_0$ by requiring that sequences of operators $\a_1, \a_2, \a_3, \dots \in \cA_0$ converge weakly to an operator $\a$ when $n \rightarrow \infty$ if their matrix elements converge 
\be
\bra{\psi} \a_n \ket{\phi} \rightarrow \bra{\psi} \a \ket{\phi}~, \quad \forall ~\ket{\psi}, \ket{\phi} \in \cH~.
\ee
This condition is called weak-operator convergence. The resulting algebra $\cA$ is a von Neumann algebra and the Hilbert space is the GNS (Gelfand–Naimark–Segal) Hilbert space. This procedure is very generic and applies to a large range of physical contexts.

We define a state $\omega$ as a linear map $\omega:\cA \rightarrow \mathbb{C}$ that satisfies the positivity condition $\omega(\a\a^\dagger) \geq 0$ for all $\a \in \cA$. The quantity $\omega(\a)$ is interpreted as the expectation value of the observable $\a$ in the state $\omega$. Let $\cH$ a Hilbert space carrying a representation $\pi$ of $\cA$ and let $\ket{\psi} \in\cH$. Then the map $\omega : \cA \rightarrow \mathbb{C}$ is given by 
\be
\omega(\a) = \bra{\psi} \pi(\a) \ket{\psi}~,
\ee
which is a state on $\cA$. Any strictly positive state (\ie, $\omega(\a \a^\dagger)>0$ for all $\a \neq0$) is called faithful. Moreover, if $\omega$ is continuous with respect to the weak-operator topology then it is called normal. Notice that this definition of state coincides with the usual definition of density matrix when the number of degrees of freedom are finite. Indeed, if the Hilbert space $\cH$ has an orthonormal basis $\{\ket{i},=1,...,n\}$ then $\omega(\ket{j}\bra{i})= \bra{i} \rho \ket{j}$~. 

A tracial von Neumann algebra is a von Neumann algebra equiped with a faithful normal tracial state $\tau$ on $\cA$. That is, $\tau$ is a linear functional $\tau : \mathcal{A} \to \mathbb{C}$ that is a faithful normal state and satisfies $\tau(\a \b)=\tau(\b \a)$. Tracial states are defined only for Type I and Type II$_1$ von Neumann algebras. With this trace at hand, we can now determine the entanglement structure of the state $\omega$ on $\cA$ by defining a density matrix operator $\rho \in \cA$
\be
\omega(\a) = \tau(\rho_\omega~\a)~,~ \forall~ \a \in \cA~.
\ee
Importantly, this approach requires only local information, as the density operator $\rho_\omega$ is tied to the state $\omega$, making explicit that the entropy is a property of the algebra $\cA$. It is a fundamental property of Type III algebras that no such density matrix exists.

A more practical way of determine von Neumann algebras is given by the bicommutant theorem of von Neumann
\begin{thm}{(von Neumann bicommutant theorem). Let $\cA \subset B(\cH)$ be a self-adjoint set. Then the commutant $\cA'$ is a von Neumann algebra and $\cA''$ is the smallest von Neumann algebra containing $\cA$. Furthermore, $\cA''=\cA$.}
\end{thm}

This theorem provides a purely algebraic characterization of a von Neumann algebra $\cA$. In many physical applications this result is extremely useful. These abstract definitions become more concrete when applied to a specific system, for instance, the algebra of observables in field theory \cite{Leutheusser:2021frk, Witten:2021jzq, Gesteau:2024dhj, Jensen:2023yxy, Penington:2023dql, Witten:2023xze, Faulkner:2024gst, Chandrasekaran:2022eqq} and cosmology \cite{Chandrasekaran:2022cip, Speranza:2025joj, Chen:2024rpx, Espindola:2026ekv, Chen:2025tbh}.
 
\subsubsection{Tomita-Takesaki theory}
\label{sec:Tomita}
The main motivation for studying von Neumann algebras is Tomita-Takesaki theory \cite{Takesaki:1970aki}. In recent years this powerful framework has led to important advances across large areas of physics and mathematics. Here we review only the basic elements that will be needed later in the paper.

Let $\cA$ be a von Neumann algebra on a Hilbert space $\cH$ that contains a cyclic and separating vector $\ket{\psi}$. The Tomita operator is defined by its action on the dense subspace generated by the algebra
\be
S_{\psi} ~ \a \ket{\psi} = \a^\dagger \ket{\psi}~ \quad \forall~\a \in \cA~.
\ee
This is an antilinear operator, which acts on a dense subset of $\cH$ and satisfies some simple properties: $S_{\psi}^2=1$, $S_\psi \ket{\psi} = \ket{\psi}$ and $S_{\psi}' = S_{\psi}^\dagger$. The Tomita operator admits a unique polar decomposition. Let $\Delta$ be the positive self-adjoint modular operator and let $J$ be the anti-unitary modular conjugation. Then
\be
S_\psi = J \Delta^{1/2} = \Delta^{-1/2} J~.
\ee
Uniqueness of the polar decomposition immediately implies $J^2=1$ and $J=J^\dagger$. 

The modular operator $\Delta$ generates a one-parameter continuous group of unitaries $\{ \Delta^{it}~\lvert~ t\in \mathbb{R} \}$. Together with the polar decomposition, this leads to the fundamental result of the theory:

\begin{thm}{(Tomita-Takesaki).}
Let $\cA$ be a von Neumann algebra with a cyclic and separating vector $\ket{\psi}$. Then $J \ket{\psi}=\ket{\psi}=\Delta \ket{\psi}$ and the following equalities hold:
\begin{equation*}
J \cA J = \cA'~, \quad {\rm and} \quad \Delta^{it} \cA \Delta^{-it} = \cA~, \quad \forall ~ t \in \mathbb{R}~.
\end{equation*}
\end{thm}
As a direct consequence, the family of unitaries $\Delta^{it}$ induces the modular automorphism group $\{ \sigma_t\}$ of $\cA$ via
\be
\sigma_t(\a) = \Delta^{it} \a \Delta^{-it}~, \quad  \a \in \cA,~ t\in \mathbb{R}~.
\ee

Given a cyclic and separating vector $\ket{\psi}$ and another vector $\ket{\phi}$, we define the relative Tomita operator $S_{\psi \lvert \phi}$ by 
\be
S_{\psi \lvert \phi} \a \ket{\psi} = \a^\dagger \ket{\phi}~.
\ee
The relative modular operator is defined by the following expression
\be
\Delta_{\psi \lvert \phi} = S^\dagger_{\psi \lvert \phi} S_{\psi \lvert \phi}~,
\ee
which is positive self-adjoint. The relative modular operator is the key object that enters the definition of relative entropy by Araki’s formula (see Sec. \ref{sec:araki}).

Finally, the modular operator provides a complete algebraic classification of von Neumann algebras. Let $\omega$ be a faithful normal state on $\mathcal{A}$ induced by the vector $\ket{\psi}$, and let $\Delta_\omega$ be the associated modular operator. The intersection
\be
S(\cA) = \bigcap_\omega~ {\rm sp}( \Delta_\omega)~,
\ee
taken over all faithful normal states $\omega$ of $\mathcal{A}$, is an algebraic invariant of $\mathcal{A}$\footnote{That is, if two algebras are isomorphic $\cA_1 \cong \cA_2$, then $S(\cA_1) = S(\cA_2)$.}. The following theorem by Connes \cite{Connes1973} classifies the possible spectra 
\begin{thm}{(Modular spectrum).}\label{tm:spectrum}
    Let $\cA$ be a factor acting on a separable Hilbert space. If $\cA$ is of Type III, then $0\in S'(\cA)$; otherwise, $S'(\cA) = \{0,1\}$ if $\cA$ is of type I$_{\infty}$ or II$_{\infty}$ and $S'{(\cA)}=\{1\}$ if not. Let $\cA$ be a factor of type III. 
    \begin{itemize}
        \item (i) $\cA$ is of type III$_{\lambda}, 0<\lambda<1$, iff $S(\cA)=\{0\}\cup \{\lambda^n~\lvert~ n \in \mathbb{Z}\}$.
        \item (ii) $\cA$ is of type III$_{0}$ iff $S(\cA)=\{0,1\}.$
        \item (iii) $\cA$ is of type III$_1$ iff $S(\cA)=[0,\infty)$.
    \end{itemize}
\end{thm}

\section{Algebraic framework} 
\label{sec:3}
In this section we develop the algebraic framework required to quantify the quantum information content carried by a Hawking pair in the centaur geometry. We begin in Sec.~\ref{sec:centaur} by introducing the centaur-algebra of observables \cite{Aguilar-Gutierrez:2023odp} $\mathcal{A}$, the Type III$_1$ von Neumann algebra generated by matter fluctuations in the right exterior of the flow geometry. In Sec.~\ref{sec:Hawkingpair} we construct the probe Hawking-pair state $\ket\Phi$ by acting with a pair of mirror operators (obtained from the modular conjugation) on $\ket\Psi$. Finally, in Sec.~\ref{sec:crossed} we promote $\mathcal{A}$ to a Type II$_\infty$ von Neumann algebra via the crossed-product construction. This enables a rigorous definition of gravitational and relative entropies, which crucially separates the gravitational and field theory contributions to the entropy difference between the reference state and the Hawking-pair state. The resulting algebraic formula contains the information exchange of the Hawking pair, which will be evaluated explicitly in later sections for flow geometries using the holographic dictionary.

\subsection{The centaur-algebra of observables}
\label{sec:centaur}
In a quantum field theory defined on a fixed spacetime, we can associate to any open spacetime region $U \subset \mathcal{M}$ a von Neumann algebra $\cA_U$ of observables. In a gravitational theory the situation is more subtle. The spacetime itself is dynamical, and large fluctuations generically prevent a sharp geometric definition of $U$. In the semiclassical limit $G_N \to 0$ (generally coupled to matter), however, metric fluctuations are parametrically suppressed and an approximate notion of a spacetime region (together with its associated algebra of observables) can be made precise. This is the regime in which we work throughout. 

\begin{figure}[t]
\begin{center}
\includegraphics[width=7.cm]{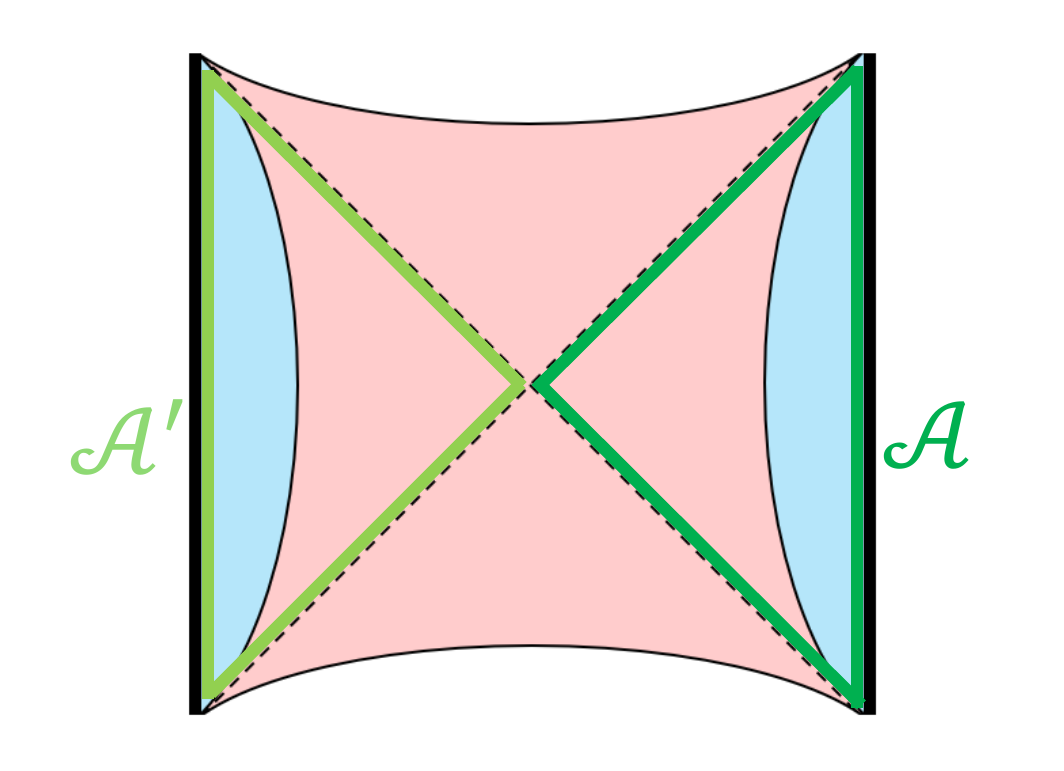} 
	\caption{Sharp centaur geometry. In the right exterior region $\mathcal{M}_{R}$, the algebra generated by bulk fluctuations around the TFD state \eqref{eq:TFD} is denoted by $\mathcal{A}$. In the left exterior region $\mathcal{M}_{L}$, its commutant is defined and denoted by $\mathcal{A}'$.
    }
\label{fig:centaur2}
\end{center}
\end{figure}

We focus on the two-dimensional flow geometries reviewed in Sec. \ref{sec:flow}. For concreteness we employ the sharp-centaur solution written in Schwarzschild gauge (Fig. \ref{fig:centaur2}, see also Fig. \ref{fig:centaur1} (Right))
\be
\label{eq:sharpL}
ds^2 =  -\left( \frac{4 \pi^2 \ell^2}{\beta^2} + \frac{1}{\ell^2}\frac{r^3}{\lvert r \lvert} \right) dt^2 + \frac{dr^2}{\left( \frac{4 \pi^2 \ell^2}{\beta^2}+\frac{1}{\ell^2}\frac{r^3}{\lvert r \lvert} \right)}~ , \quad -\frac{2\pi \ell^2}{\beta} \leq r < \infty~. 
\ee
The Lorentzian geometry is an extended two-sided spacetime that is asymptotically AdS$_2$ at both boundaries. We denote the right and left exterior regions by ${\cal{M}}_L$ and ${\cal{M}}_R$ respectively. The dynamics of the centaur theory is captured by a Schwarzian $SL(2,\mathbb{R})$-invariant boundary action supplemented with Dirichlet boundary conditions on both the dilaton and the induced metric\footnote{For a general discussion of boundary conditions in flow geometries see \cite{Anninos:2018svg}.}.

In the absence of matter the one-sided Hamiltonians $H_R$ and $H_L$ coincide, and the algebra of gauge-invariant operators $\cA$ (respectively the commutant $\cA'$) consists of operators invariant under residual $SL(2,\mathbb{R})$ transformations generated by $\Phi_R(t)$. This algebra is commutative and satisfies $\cA \cap \cA' \neq 0$, reflecting the fact that the two exteriors share a common asymptotic boundary. The apparent factorization problem is resolved once matter is included.

We now couple the geometry to matter fields. Working in the generalized free field approximation, matter fields admit the decomposition in Fourier modes with respect to the boundary time $u$ 
\be
\phi (u) = \int \frac{d \omega d\lambda}{4 \pi^2} \left(f_{\omega}(u) a_{\omega} + f_\omega^*(u) a^\dagger_{\omega} \right)~.
\ee
Let $\cA$ be the von Neumann algebra generated by bulk fluctuations $\{ a, a^\dagger \}$ associated with the right exterior region $\mathcal{M}_R$. A generic element $\a \in \cA$ is obtained as the weak-operator limit of finite linear combinations of these modes. By the HKLL reconstruction, these operators are identified with (light) operators at the AdS$_2$ boundary; equivalently, they are operators in the flowed CFT dual to the centaur geometry. In the presence of dynamical matter the left and right Hamiltonians no longer coincide, commutativity is lost, and $\mathcal{A}$ (as well as its left counterpart) becomes a Type III$_1$ von Neumann algebra.

Having specified the local algebra of observables, we must choose the class of states on which it acts. We are interested in the thermofield double (TFD) state of the left and right boundary theories at inverse temperature $\beta$ (see Fig. \ref{fig:centaur2})
\be
|\Psi\rangle = \frac{1}{\sqrt{Z_\beta}} \sum_a e^{-\beta E_a/2} |E_a\rangle_L |E_a\rangle_R~, \label{eq:TFD}
\ee
where $Z_\beta$ is the canonical partition function. This state is the natural thermal purification compatible with the Euclidean periodicity that defines the centaur background. We will use the microcanonical version later in Sec. \ref{sec:Sgen} when we describe the generalized entropy in the Hawking pair framework. Acting with elements $\a \in \mathcal{A}$ on $\ket{\Psi}$ generates the GNS Hilbert space $\mathcal{H}_{\rm Flow}$, which is dense in the full physical Hilbert space. 

\subsection{Hawking pair state}
\label{sec:Hawkingpair}
We now construct the probe state that keeps track of the quantum information content carried by a single Hawking pair as a function of the Euclidean-time separation.

Start with a local HKLL operator $\phi_0 \in \cA$ inserted at the AdS$_2$ boundary at Euclidean time $\tau=0$. We can then evolve it in Euclidean time using the modular Hamiltonian $h_\Psi$ associated with the algebraic state $\omega$ induced by the TFD state $\ket\Psi$ 
\be
\phi_\tau = e^{- \tau h_\Psi} \phi_0 e^{\tau h_\Psi}~, ~\phi_\tau \in \cA~,
\ee
where Euclidean time runs $\tau \in (0, \beta/4)$. This operator creates a bulk excitation that propagates toward the $t=0$ slice. In the large-mass limit the excitation localizes along a classical trajectory. From the point of view of a bulk observer in the state $\Psi$, this represents a Hawking-like state.

The Hawking partner of $\phi_\tau$ is the mirror operator $\bar{\phi}_\tau$ living in the commutant algebra $\cA'$. It is obtained from the Tomita operator $S_\Psi$ on the state $\ket\Psi$ via the modular conjugation $J_\Psi$
\be
\ol{\phi}_\tau = J_\Psi \phi_\tau J_\Psi~.
\ee
By Tomita–Takesaki theory (see Sec. \ref{sec:Tomita}), $J_\Psi$ implements the duality $J_\Psi \mathcal{A} J_\Psi = \mathcal{A}'$, so $\bar{\phi}_\tau$ indeed lies in the commutant. Moreover, because $J_\Psi$ reverses modular flow we have the relation
\be
J_\Psi \phi_\tau J_\Psi = \phi_{\tilde{\tau}} \quad \text{with} \quad \tilde{\tau} = \frac{\beta}{2} - \tau~,
\ee
placing the mirror partner precisely behind the horizon on the $t=0$ slice. Acting with the mirror operator on $\ket{\Psi}$ likewise produces a bulk excitation. 

We define the Hawking-pair state by acting with both operators on the TFD state
\begin{figure}[t]
\begin{center}
\includegraphics[width=13.cm]{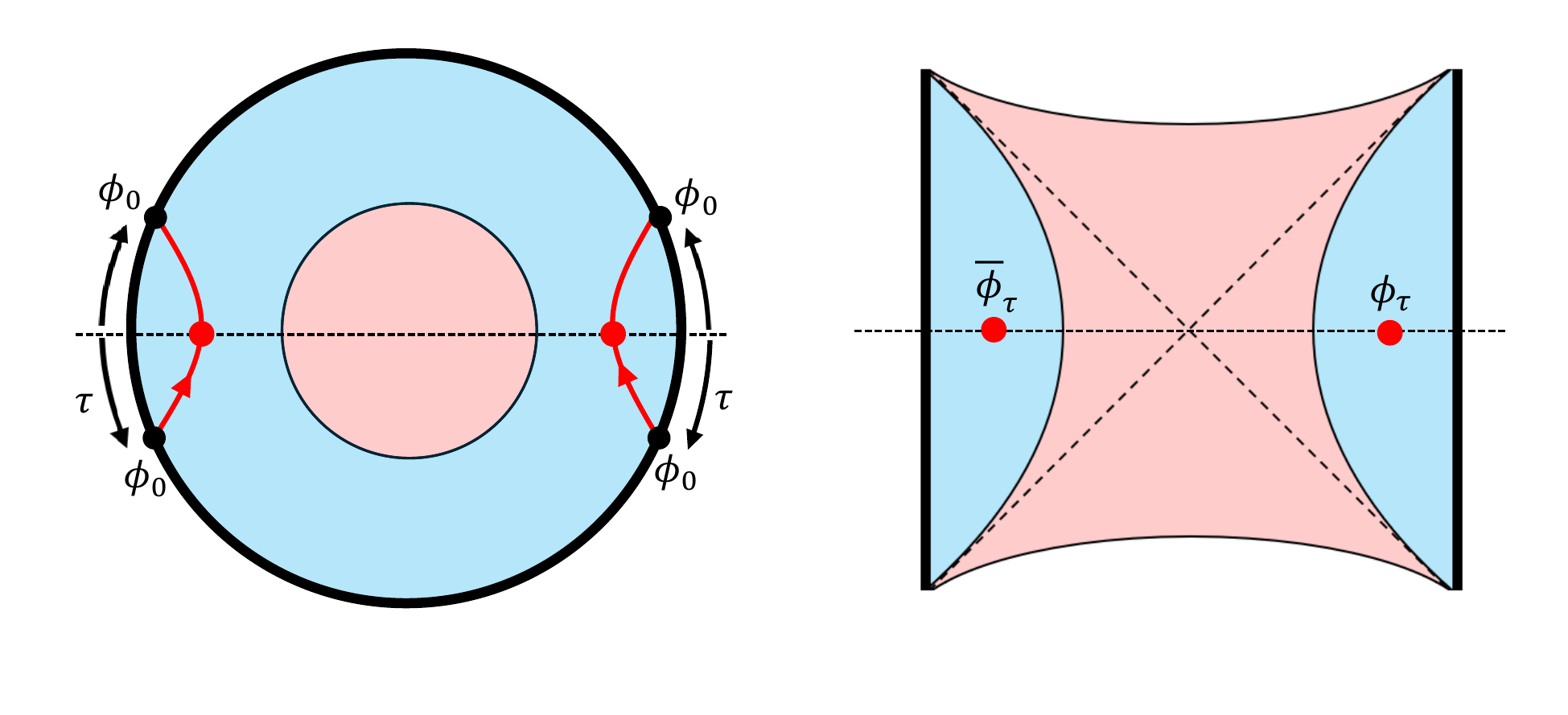} 
	\caption{(Left) Euclidean preparation of the Hawking pair state \eqref{eq:HPstate}. The right trajectory is created by inserting the operator $\phi_{0}$ on the boundary at $\tau$ and $-\tau$, while the left trajectory is created by inserting it on the boundary at $\frac{\beta}{2}-\tau$ and $\frac{\beta}{2}+\tau$.  (Right) The Euclidean preparation corresponds to acting with the bulk local operators $\phi_{\tau}$ and $\overline{\phi}_{\tau}$ on the TFD state. We identify these operators with a Hawking pair. Their locations are given by the intersections of the Euclidean trajectories with the time-symmetric slice.
    }
\label{fig:HPstate}
\end{center}
\end{figure}
\be
\ket \Phi := N^{-1/2} \phi_\tau \bar{\phi}_\tau \ket \Psi~. \label{eq:HPstate}
\ee
Here $N$ is the normalization factor proportional to the thermal four point function. The mirror operator $\bar{\phi}_\tau$ is located behind the horizon at the mirror location $\tilde{\tau}= \beta/2 - \tau$. The state is depicted in Fig. \ref{fig:HPstate}.

Having constructed the Hawking-pair probe state directly on the centaur-algebra $\mathcal{A}$, we now have a microscopic operator that creates an entangled pair of conjugate quanta, one outside and one behind the cosmological horizon. To extract the quantum information content carried by this pair we must first promote the algebra to a Type II$_\infty$ factor by dressing the operators with the Hamiltonian. This is accomplished by the crossed-product construction, which we present in the next subsection. The resulting Type II$_\infty$ algebra supplies a well-defined trace and a clean definition of entropies together with the entropy difference between states $\ket\Psi$ and $\ket\Phi$.

\subsection{Crossed-product algebra}
\label{sec:crossed}

To compute the information content carried by a single Hawking-pair in the centaur geometry, we must dress the centaur-algebra $\mathcal{A}$ with the gravitational degree of freedom. This is accomplished via the crossed-product construction, which enlarges $\mathcal{A}$ to a Type II$_\infty$ von Neumann algebra 
\be
\hat{\mathcal{A}} = \mathcal{A} \rtimes_{\sigma^\Psi_t} \mathbb{R}~,
\ee
where $\sigma^\Psi_t$ is the modular automorphism group of any faithful reference state $\ket{\Psi}$ on $\mathcal{A}$. The crossed product preserves all correlation functions of operators in $\mathcal{A}$, while eliminating the center that appears in the strict large-$N$ limit. 

Let $X$ be a self-adjoint generator of the unitary representation of $\mathbb{R}$ on $L^2(\mathbb{R})$, $U(t) = e^{itX}$. This generator satisfies the relation 
\be
e^{itX} \a \, e^{-itX} = \sigma^\Psi_t(\a)~, \qquad \a\in\mathcal{A}~.
\ee
We adjoin all bounded functions of $h_\Psi +X$ to the algebra $\mathcal{A}$. A generic element of the crossed-product algebra takes the form
\be
\hat{\a} = \a \, e^{i\alpha (h_\Psi + X)}~, \qquad \a \in \mathcal{A}~, \quad \alpha \in \mathbb{R}~,
\ee
where $h_\Psi = -\log\Delta_\Psi  $ is the modular Hamiltonian, which satisfies $h_\Psi|\Psi\rangle=0$. The algebra $\hat{\mathcal{A}}$ acts on the extended Hilbert space 
\be
\hat{\mathcal{H}}_{\rm Flow} = \mathcal{H}_{\rm Flow} \otimes L^2(\mathbb{R})~,
\ee
with the second factor carrying the auxiliary classical time mode generated by $X$. The semiclassical reference state on this enlarged space is the classical-quantum product state
\be
\ket{\hat{\Psi}} := \ket \Psi \otimes f(X)~,
\ee
with $f(X)\in L^2(\mathbb{R})$ a normalized wavefunction. The modular operator $\hat{\Delta}_{\hat{\Psi}}$ on $\hat{\mathcal{A}}_R$ satisfies the relation
\be
\langle \hat{\Psi} | \hat{\a}~\hat{\b} | \hat{\Psi} \rangle = \langle \hat{\Psi} | \hat{\b} \, \hat{\Delta}_{\hat{\Psi}} \, \hat{\a} | \hat{\Psi} \rangle ~, \qquad \hat{\Delta}_{\hat{\Psi}} = \hat{K}\hat{K}^\dagger~,
\ee
where the operator
\be
\hat{K} = e^{-(h_\Psi + X)} f(h_\Psi + X)~,
\ee
in $\hat{\mathcal{A}}$, defines the trace on the crossed-product algebra
\be
\label{eq:trace}
{\rm Tr} (\hat{\a}) = \int_{-\infty}^\infty dX ~ e^X \bra \Psi  \a(X) \ket \Psi ~.
\ee
See Appendix \ref{App:crossed} for a review of this result. Identifying the right-moving generator via the standard dictionary $h_R := (h_\Psi + X)/2$, with the modular flow normalized so that its generator matches the Hawking temperature $\beta$, the density matrix of $\ket{\hat{\Psi}}$ on $\hat{\mathcal{A}}  $ is
\be\label{densitypsi}
\rho_{\hat{\Psi}} = |f(h_R)|^2 \, e^{-\beta h_R}~.
\ee
The entropy of the density matrix is
\be 
S(\rho_{\hat{\Psi}}) = \int_{-\infty}^\infty dX \bigl( X g(X) - g(X) \log g(X) \bigr)~,
\ee
with $g(X) \in L^2(\mathbb{R})$. The trace on $\hat{\mathcal{A}}$ is not canonical, as the algebra admits a group of outer automorphisms that shifts $X \to X + c$ with $c \in \mathbb{R}$. Under this shift the trace rescales as $  \operatorname{Tr} \to e^c \operatorname{Tr}$. To preserve the normalization condition $\operatorname{Tr} \hat{\rho} = 1  $, one must compensate by rescaling the density matrix $  \hat{\rho} \to e^{-c} \hat{\rho}$. This has the effect of shifting the entropy by the same constant $S(\hat{\rho}) \to S(\hat{\rho}) + c.$

The relative modular operator $\Delta_{\Phi|\Psi}$ between two faithful normal states $\ket{\Phi}$ and $\ket{\Psi}$ on $\hat{\mathcal{A}}$ satisfies 
\be
\log\Delta_{\Phi|\Psi} = \log\rho_{\hat{\Phi}} - \log\rho_{\hat{\Psi}}~.
\ee
For the perturbed state $\ket \Phi$ the corresponding density matrix is 
\be\label{densityphi}
\rho_{\hat{\Phi}} = |f(h_R)|^2 \, e^{-\beta(h_R - h_\Psi + h_{\Phi|\Psi})}~.
\ee
This form follows directly from the Connes cocycle together with the fact that both states are thermal with respect to the same fixed inverse temperature $\beta$. 

We can compute the entropy difference $\Delta S$ between the state with and without the Hawking pair. The algebraic approach allows us to clearly distinguish between the gravitational and QFT contributions. Using the density matrices \eqref{densitypsi} and \eqref{densityphi} we get
\begin{align}
    \Delta S = {\rm Tr~}{\rho_{\hat{\Phi}}} \log\rho_{\hat{\Phi}} - {\rm Tr~}{\rho_{\hat{\Psi}}} \log\rho_{\hat{\Psi}} 
    = \Delta S_{\rm grav} (\Phi, \Psi) - \Delta S_{\rm rel}(\Phi \lvert \lvert \Psi)~,
\end{align}
where we have defined\footnote{ \label{footnote:Srel}Here, $\Delta S_{\rm rel}$ denotes the usual relative entropy for a Type III algebra. The unusual notation with $\Delta$ is intended to emphasize that it represents the QFT contribution to the entropy difference $\Delta S$. We have dropped the hats from the states in $\Delta S_{\rm rel}(\Phi \lvert\lvert \Psi)$ to emphasize that this quantity is associated with the QFT algebra.}
\be
\Delta S_{\rm grav} (\Phi, \Psi) := \bra{\hat{\Phi}} \beta h_R \ket{\hat{\Phi}}~, \quad \Delta S_{\rm rel}(\Phi \lvert \lvert \Psi) := \bra \Phi h_{\Psi | \Phi} \ket \Phi~.
\ee
Notice that the states appearing in $\Delta S_{\rm grav}$ refer to the full classical-quantum product state on the extended Hilbert space. This quantity has a characteristic behavior that can be used to determine the amount of information that a Hawking pair extracts from black holes \cite{Verlinde:2022xkw}. Our main goal is to compute the information exchange of this Hawking pair to the exterior, thus extracting information from the dS horizon. These algebraic manipulations boil down to computing correlation functions between Hawking partners. In the next section, we make use of the holographic dictionary to compute the boundary Euclidean two-point functions $G_\beta(\Delta\tau)$ that enter both in the normalization $N$ and the Wick contractions required to compute the entropy change.

\section{Euclidean boundary two-point function of centaur gravity} \label{sec:twopoint}
In this section we compute the Euclidean boundary two-point function in the centaur geometry. We begin in Sec. \ref{subsec:geoAdS} by analyzing geodesics that remain entirely within the AdS region. In Sec. \ref{subsec:geodS} we study geodesics that enter the dS hemisphere. Finally, in Sec. \ref{subsec:twoptfunc} we use the geodesic approximation in the large-mass limit to evaluate the boundary two-point function, showing that the dominant contributions come from trajectories that stay within the AdS region. The resulting correlator serves as a key input for the entropy calculations performed in Sec. \ref{sec:5}.

\begin{figure}[t]
\begin{center}
\includegraphics[width=12.cm]{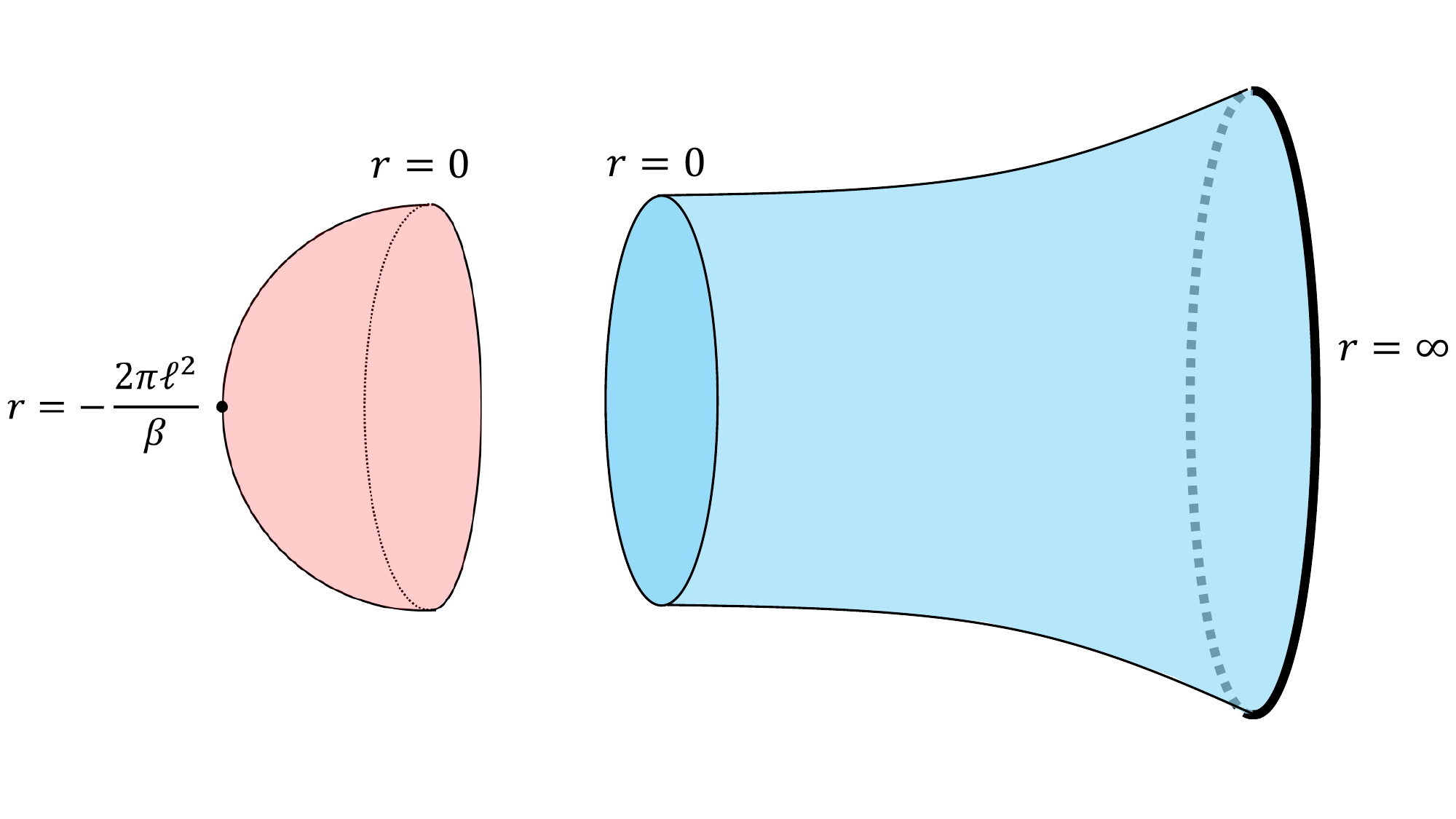} 
	\caption{(Left) Hemisphere. In the coordinate \eqref{eq:dShemisphere}, $r=0$ represents 
    the equator and $r=- \frac{2\pi \ell^2}{\beta}$ represents a pole. (Right) ``AdS-trumpet." In the coordinate \eqref{eq:AdStrumpet}, $r=\infty$ corresponds to the AdS boundary.
    }
\label{fig:AdSdS}
\end{center}
\end{figure}

As reviewed in Subsection \ref{sec:flow}, the centaur geometry is constructed by gluing the AdS region and the dS region. In the Euclidean signature, the AdS region is an``AdS-trumpet" (\ie a half of the double trumpet)
\be
\begin{array}{c}
\displaystyle \rmd s^2_{\rm AdS-trumpet} = \left( \frac{4\pi^2 \ell^2}{\beta^2} + \frac{r^2}{\ell^2} \right) d\tau^2 + \frac{1}{\left( \frac{4\pi^2 \ell^2}{\beta^2} + \frac{r^2}{\ell^2} \right)} dr^2  \\
\tau \in [0,\beta], ~~~ r\in [0, \infty)
\end{array}
\hspace{1cm} ({\rm Fig.~ \ref{fig:AdSdS} ~ (Right)}) ~, \label{eq:AdStrumpet}
\ee
and the dS region is a hemisphere
\be
\begin{array}{c}
\displaystyle \rmd s^2_{\rm dS-hemisphere} = \left( \frac{4\pi^2 \ell^2}{\beta^2} - \frac{r^2}{\ell^2} \right) d\tau^2 + \frac{1}{\left( \frac{4\pi^2 \ell^2}{\beta^2} - \frac{r^2}{\ell^2} \right)} dr^2  \\
\tau \in [0,\beta], ~~~ r\in [-\frac{2\pi \ell^2}{\beta}, 0]
\end{array}
\hspace{1cm} ({\rm Fig.~ \ref{fig:AdSdS}~ (Left)}) ~ . \label{eq:dShemisphere}
\ee
Here, $\beta$ represents the inverse temperature of the horizon and these geometries are glued at $r=0$ to form the flow geometry \eqref{eq:sharpE}.

As is well-known, for any two points on the AdS boundary $r=\infty$ of the AdS-trumpet, we can always find geodesics which connects them staying in this geometry, \ie, ones do not cross the $r=0$ surface. This type of geodesics is analyzed in Subsection \ref{subsec:geoAdS}. On the other hand, we can also find geodesics which connect a point on the AdS boundary and one on $r=0$ surface, thus boundary-to-boundary geodesics entering the dS region. This type of geodesics is analyzed in \ref{subsec:geodS}.

As we will soon see in these subsections, we have infinitely many geodesics for given two boundary points in the centaur geometry. In Subsection \ref{subsec:twoptfunc}, for the purpose to use them for the geodesic approximation of the two-point function of heavy mass, we will show that only the geodesics staying AdS region (\ie, does not entering the dS region and ones analyzed in Subsection \ref{subsec:geoAdS}) are relevant.

\subsection{Geodesics in the AdS region} \label{subsec:geoAdS}
\begin{figure}[t]
\begin{center}
\includegraphics[width=14.cm]{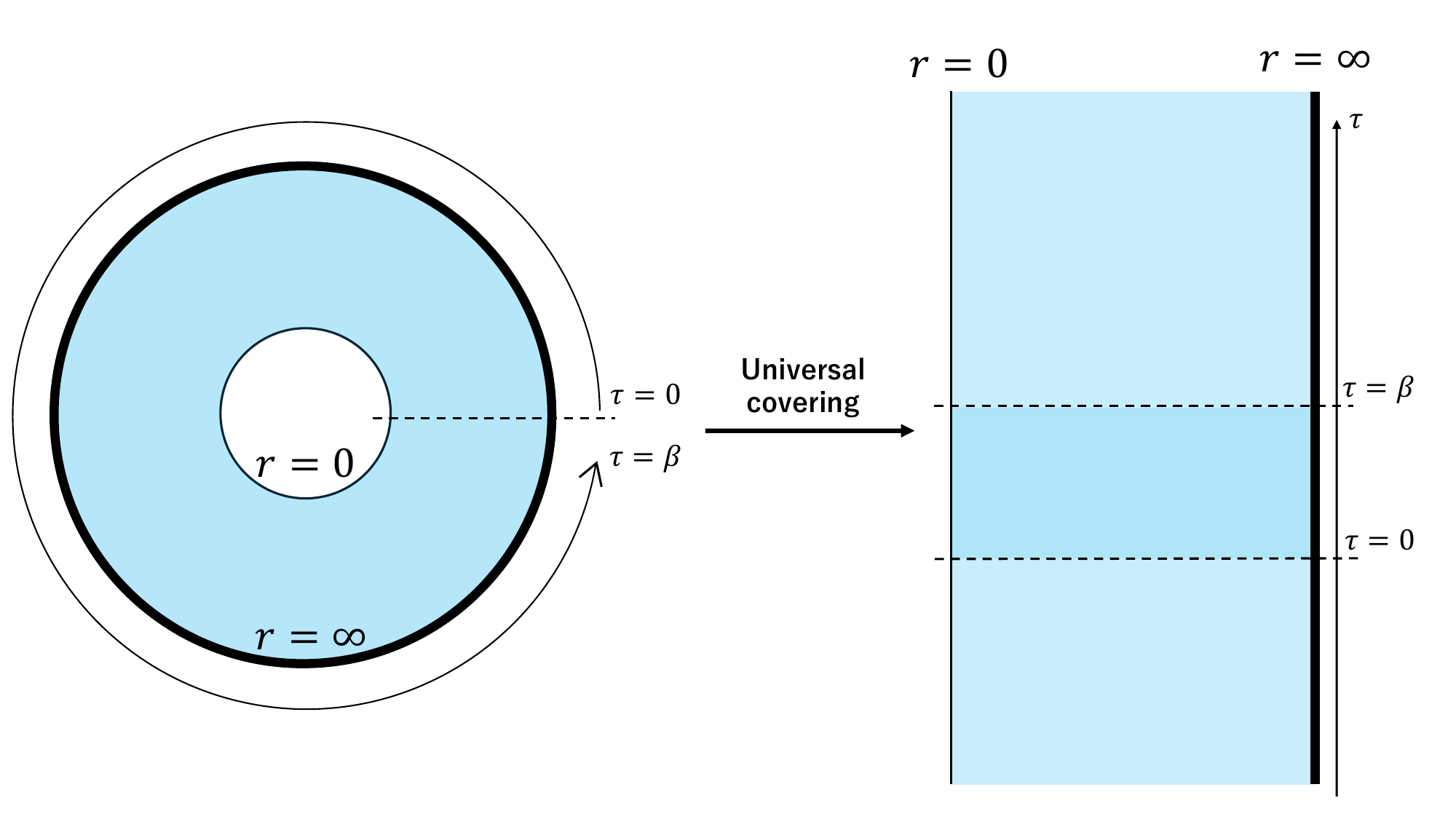} 
	\caption{(Left) AdS-trumpet. (Right) Universal cover of the AdS-trumpet. The range of $\tau$ is extended to $(-\infty, +\infty)$.
    }
\label{fig:UC}
\end{center}
\end{figure}

To understand the behavior of geodesics in the AdS-trumpet, it is presumably instructive to begin with that in its universal cover by changing the range of $\tau$ to $(-\infty, \infty)$ (see Fig. \ref{fig:UC}).
Let $f_{AdS}(r)$ be the metric function in the AdS-trumpet \eqref{eq:AdStrumpet}:
\be
f_{AdS}(r) = \left( \frac{4\pi^2 \ell^2}{\beta^2} + \frac{r^2}{\ell^2} \right) ~ .
\ee
To avoid introducing unnecessary characters, let us denote the geodesic by $(\tau,r)=(\tau(s),r(s))$ with a parameter $s$. Although we use the same characters as the coordinate variable, we will always write them with an argument when we mean geodesics. Since we are considering a special case where the background has a time translation symmetry, there is a conserved charge and let us call it $\mathcal{E}$ and defined by
\be
\mathcal{E} = f(r(s))\dot{\tau}(s) ~ . \label{eq:geocon}
\ee
And, in general, the geodesic action has a reparametrization gauge redundancy and we fix this by demanding the parameter $s$ to be the proper time, meaning that the tangent vector is normalized to have an unit norm:
\be
f(r(s)) \dot{\tau}(s)^2 + \frac{1}{f(r(s))} \dot{r}(s)^2 = 1 ~ . \label{eq:geofix}
\ee
Therefore, in this case, we can obtain the solution without considering the EOM directly but from the combination of \eqref{eq:geocon} and \eqref{eq:geofix}:
\begin{align}
\dot{r}(s)^2 = -\mathcal{E}^2 + \frac{4 \pi^2 \ell^2}{\beta^2} + \frac{r(s)^2}{\ell^2} ~ . \label{eq:AdSrEOM} \\ 
\Longrightarrow ~~ r(s) = \ell \sqrt{\mathcal{E}^2 - \frac{4\pi^2 \ell^2}{\beta^2}} \cosh \left( \frac{s}{\ell} \right) ~ , \label{eq:solr1}
\end{align}
and from \eqref{eq:geofix}
\footnote{
Here, since the boundary condition for the geodesic at $s=\pm \infty$ respects the $\mathbb{Z}_{2}$ symmetry of the background space, we know that the solution also has a $\mathbb{Z}_{2}$ symmetry. After the gauge fixing \eqref{eq:geofix}, the geodesic theory still has a translation symmetry $s \to s+a$. We fix this freedom by requiring that $s=0$ is the fixed point of the aforementioned $\mathbb{Z}_{2}$ symmetry, \ie $(\tau(s), r(s)) = (\tau(-s), r(-s))$.
}
\be
\tau(s) = \frac{\beta}{2\pi} {\rm arctanh} \left[ \frac{2\pi \ell}{\mathcal{E} \beta} \tanh \left( \frac{s}{\ell} \right) \right] ~ . \label{eq:soltau1}
\ee 
Up to now, the solution is parametrized by its conserved charge $\mathcal{E}$. For later convenience, let us rewrite it with the boundary separation
\be
\Delta \tau = \tau(\infty) - \tau(-\infty) ~ .
\ee
Using the explicit form of the solution \eqref{eq:soltau1}, $\Delta \tau$ can be written as
\be
 \Delta \tau = \frac{\beta}{\pi} {\rm arctanh} \left[ \frac{2\pi \ell}{\mathcal{E} \beta} \right]  ~,
\ee
and thus 
\be
\mathcal{E} = \frac{2\pi \ell}{\beta} \frac{1}{\tanh (\frac{\pi \Delta \tau}{\beta})} ~ .
\ee
Substituting this into the solution \eqref{eq:solr1}, \eqref{eq:soltau1}, we obtain
\begin{align}
r(s) = \frac{2\pi \ell^2}{\beta} \frac{1}{\sinh(\frac{\pi \Delta \tau}{\beta})} \cosh(\frac{s}{\ell}) ~ , \label{eq:solr2} \\
\tau(s) = \frac{\beta}{2 \pi} {\rm arctanh} \left[ \tanh(\frac{\pi \Delta \tau}{\beta}) \tanh(\frac{s}{\ell}) \right] ~ . \label{eq:soltau2}
\end{align}
\begin{figure}[t]
\begin{center}
\includegraphics[width=7.cm]{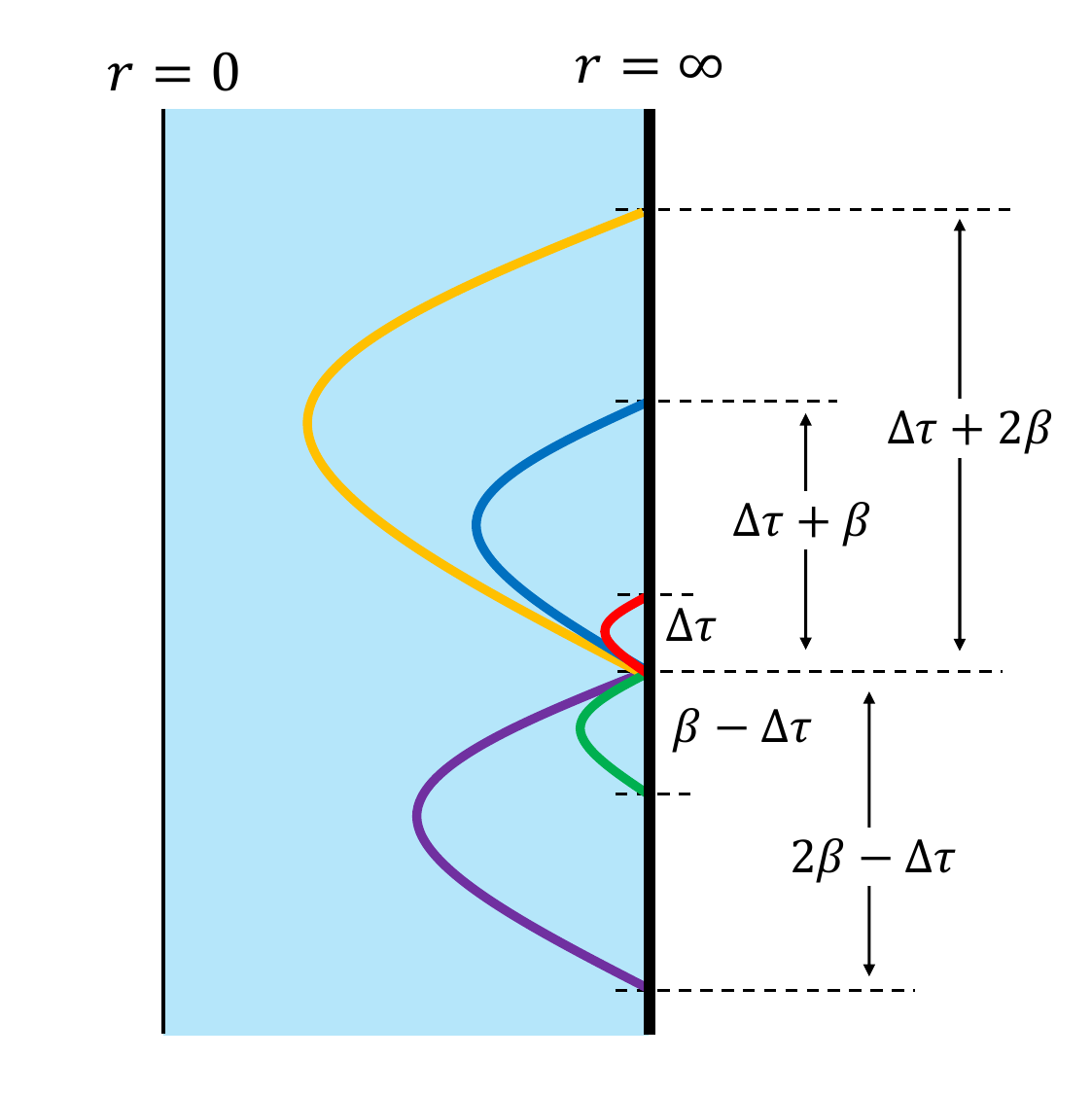} ~~\includegraphics[width=7.cm]{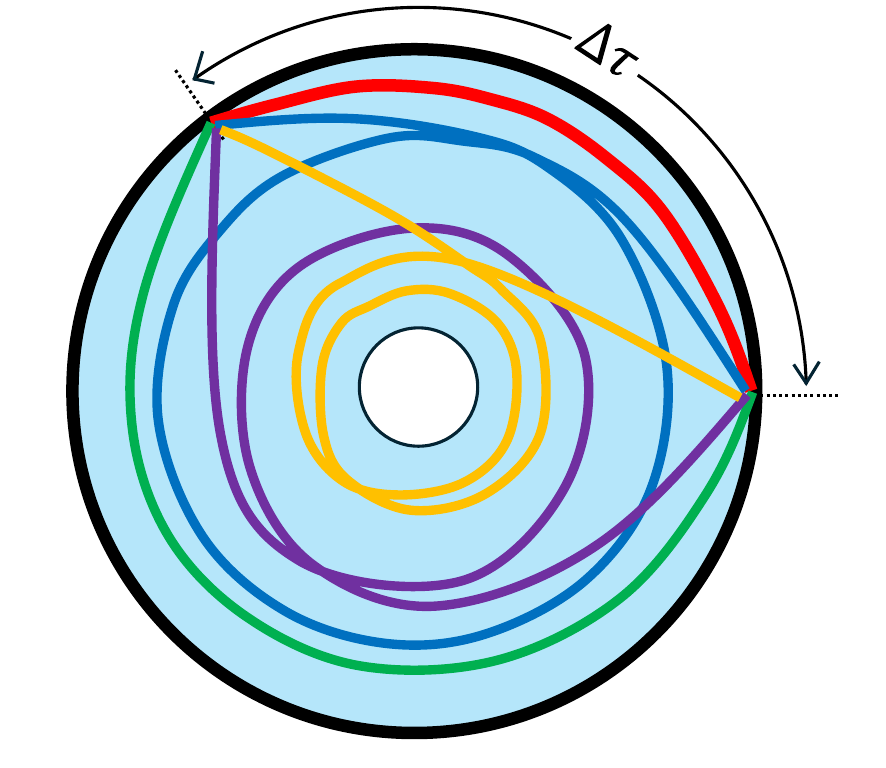} 
	\caption{(Left) Family of geodesics with different boundary separations on the universal cover. Here, we fix $\Delta \tau < \beta/2$ and $\beta$, and drew  geodesics with the separations $|\Delta \tau + n \beta|$ with $ n=-2,-1,0,1, {\rm and} ~ 2$. (Right) Corresponding family of geodesics on the AdS-trumpet. They all have the same boundary separation $\Delta \tau$. 
    }
\label{fig:geodesics}
\end{center}
\end{figure}
Fig. \ref{fig:geodesics} (Left) is a schematic picture of the family of solution with different values of $\Delta \tau$. It may be clear that, for a given value of $\Delta \tau$, there exists an unique solution and as $\Delta \tau$ increases, the geodesic goes deeper in the bulk whose minimum radius is given by
\be
r_{min} = r(0) = \frac{2\pi \ell^2}{\beta} \frac{1}{\sinh(\frac{\pi \Delta \tau}{\beta})} ~ ,
\ee
but never crosses the $r=0$ surface.

Having this behavior of geodesics in mind, let us go back to the original AdS-trumpet geometry, which is obtained by compactifying the geometry of Fig. \ref{fig:geodesics} (Left). Due to this compactification, the geodesics with the boundary separation $\Delta \tau^{({\rm universal ~ cover})}$ are identified with those of the boundary separation $\Delta \tau^{({\rm trumpet})} \leq \beta/2 $ satisfying
\begin{align}
|\Delta \tau^{({\rm trumpet})} + n \beta| = \Delta \tau^{({\rm universal ~ cover})} ~~~~~~ (n\in \mathbb{Z}) ~ . \label{eq:deltaurel}
\end{align}
See Fig. \ref{fig:geodesics} (Right). Therefore, there are an infinite number of geodesics for a given boundary separation $\Delta \tau$ in the AdS-trumpet geometry. Note that the shortest geodesic is just that of $n=0$. It may also be clear intuitively but we can confirm it by using the explicit form of the geodesic length
\be
({\rm geodesic ~ length}) \propto \log \left( \sinh\left( \frac{\pi \Delta \tau}{\beta} \right) \right) ~ , \notag
\ee
where we omit an irrelevant factor and its exact form will be derived in Subsection \ref{subsec:twoptfunc} and shown in \eqref{eq:OSactionADS}. The above formula can be applied to all geodesics for the universal covering geometry
\footnote{
For the AdS-trumpet, the above formula can only be applied to the shortest geodesics. To apply it to other geodesics, $\Delta \tau$ in the formula must be replaced appropriately using \eqref{eq:deltaurel}.
}.
It means that the shorter the boundary separation is, the shorter its geodesic length is.

\subsection{Geodesics entering the dS region} \label{subsec:geodS}

In the AdS-trumpet, or in its universal cover, there also exist geodesics which connect the boundary $r=\infty$ and the $r=0$ surface. This means that, in the full centaur geometry, they will enter the dS region $r<0$ and exit, and finally go to the boundary again, forming a boundary-to-boundary geodesic. Thus, this type of geodesic can be separated into three parts; two segments in the AdS-trumpet and one segment in the dS-hemisphere. Let us discuss them separately and briefly mention about the behavior of the full geodesic finally.
\subsubsection*{Geodesic in AdS-trumpet}
Similarly as in the previous subsection, it may be convenient to start from the universal cover of the AdS-trumpet. The equation for $r(s)$ is the same as \eqref{eq:AdSrEOM}. In the previous case, we obtained the cosh function \eqref{eq:solr1} due to its boundary condition $r(+\infty) = r(-\infty)= \infty$. Here, let us take a different boundary condition, say,
\be
r(\infty) = \infty ~, \quad r(0) = 0 ~ . \notag
\ee
This boundary condition leads to the sinh function;
\be
r(s) = \ell \sqrt{ \frac{4\pi^2 \ell^2}{\beta^2} - \mathcal{E}^2 } \sinh \left( \frac{s}{\ell} \right), ~~~ s\in [0, +\infty) ~ .
\ee
$\tau(s)$ is given by
\be
\tau(s) = \frac{\beta}{2\pi} {\rm arctanh} \left[ \frac{\mathcal{E}\beta}{2\pi \ell} \tanh \left( \frac{s}{\ell} \right) \right] ~ .
\ee
\begin{figure}[t]
\begin{center}
\includegraphics[width=5.cm]{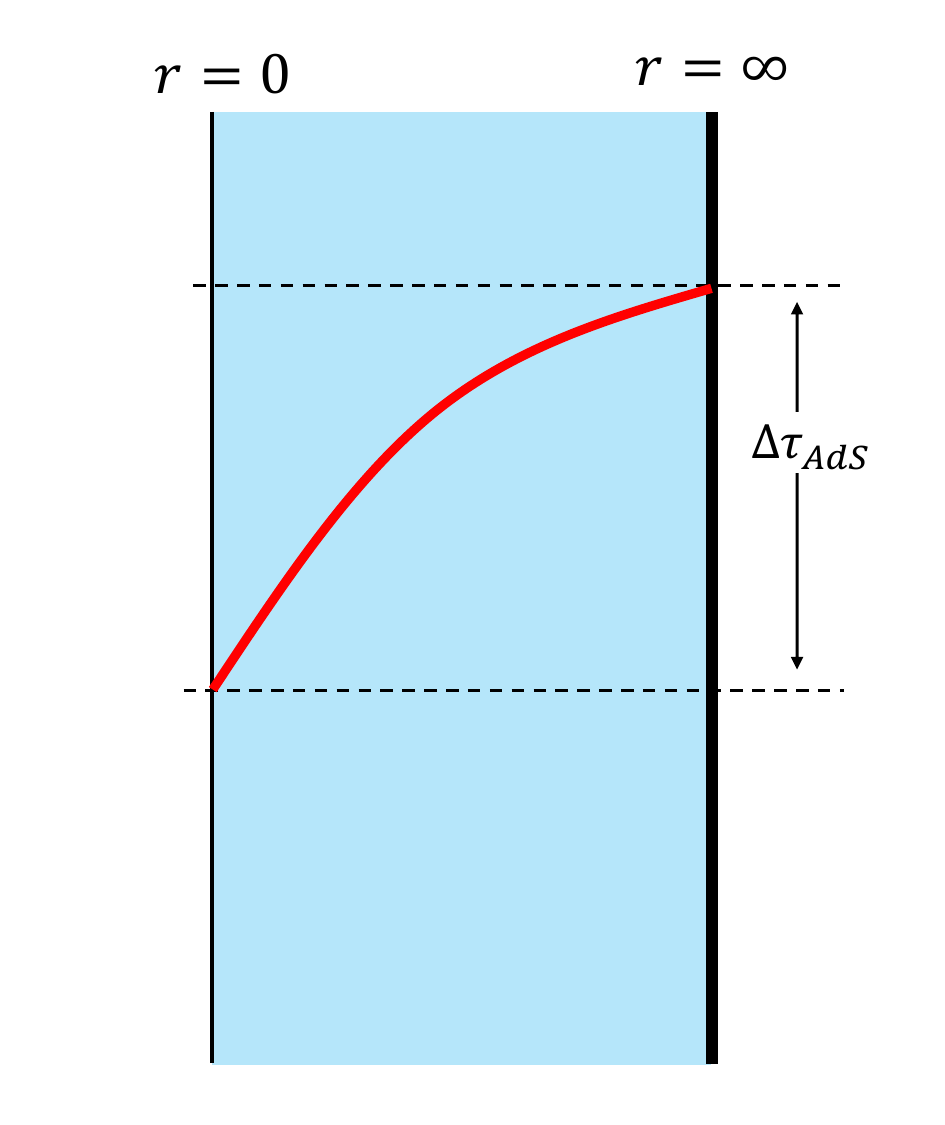} 
	\caption{On the universal cover, $\Delta \tau_{AdS}$ is defined by the difference of the Euclidean time $\tau$ at $r=0$ and $r=\infty$.
    }
\label{fig:geodesicAdS}
\end{center}
\end{figure}
Again, let us replace $\mathcal{E}$ with a Euclidean time separation. In this case, we define the time separation, denoted $\Delta \tau_{AdS}$, by $
\Delta \tau_{AdS} = \tau(\infty) - \tau(0) $ (see Fig. \ref{fig:geodesicAdS}) and given by
\begin{align}
\Delta \tau_{AdS} = \tau(\infty) - \tau(0) \hspace{1.25cm} \notag \\
= \frac{\beta}{2\pi} {\rm arctanh} \left[ \frac{\mathcal{E}\beta}{2\pi \ell} \right] ~ ,
\end{align}
and thus
\be
\mathcal{E} = \frac{2\pi \ell}{\beta} \tanh \left( \frac{2\pi \Delta \tau_{AdS}}{\beta} \right) ~ .
\ee
Then, the solution can be rewritten as
\begin{align}
r(s) = \frac{2\pi \ell^2}{\beta} \frac{1}{\cosh \left( \frac{2\pi \Delta \tau_{AdS}}{\beta} \right)} \sinh\left( \frac{s}{\ell} \right) ~, \hspace{2.05cm} \label{eq:rofs} \\
\tau(s) = \frac{\beta}{2 \pi} {\rm arctanh} \left[ \tanh\left( \frac{2\pi \Delta \tau_{AdS}}{\beta} \right) \tanh\left( \frac{s}{\ell}\right) \right] ~ . \label{eq:tauofs}
\end{align}
To understand how it enters the dS region its derivatives at $r=0$ would be useful;
\begin{align}
\dot{r}(0) = \frac{2\pi \ell^2}{\beta} \frac{1}{\cosh \left( \frac{2\pi \Delta \tau_{AdS}}{\beta} \right)} ~, \hspace{0.4cm} \\
\dot{\tau}(0) = \frac{\beta}{2\pi \ell} \tanh \left( \frac{2 \pi \Delta \tau_{AdS}}{\beta} \right) ~ . \label{eq:taudot}
\end{align}

\subsubsection*{Geodesic in the dS-hemisphere}
\begin{figure}[t]
\begin{center}
\includegraphics[width=8.cm]{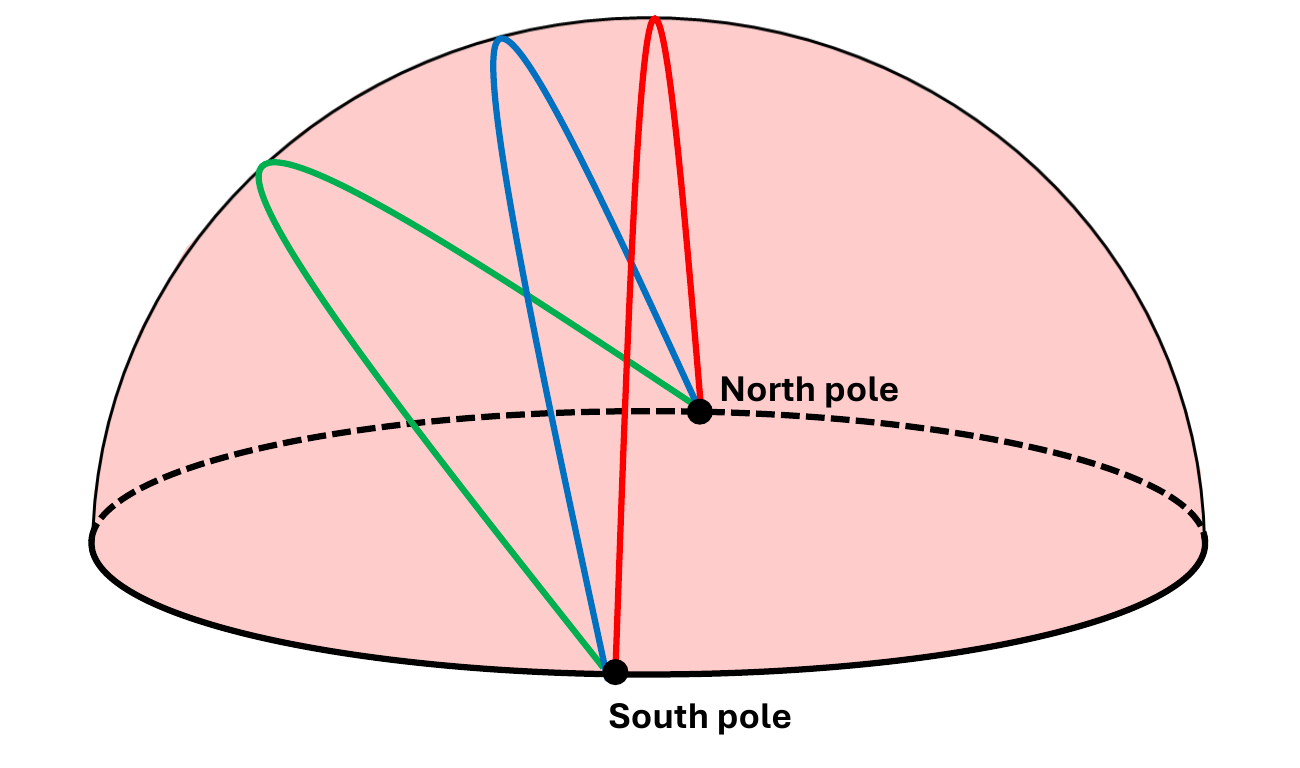}
\caption{Geodesics on the hemisphere. The inclination depends on the value of $|\dot{\tau}(0)|$ as defined in \eqref{eq:taudot}.}
\label{fig:geodesicEdS}
\end{center}
\end{figure}
A basic geometric fact about the (hemi)sphere is that geodesics always lie on great circles. Thus, on a hemisphere, the endpoints of a geodesic lie on antipodal points, say, the north and south poles. For fixed poles, which great circle the geodesic follows depends on the direction in which it is emitted from one of the poles. The inclination of the geodesic is quantified by the $\tau$ component of the velocity vector \eqref{eq:taudot} (see Fig.~\ref{fig:geodesicEdS}). The detailed form of the geodesic will not be important for our purposes. The key point is that the geodesic is always a half of a great circle, and its length is given by
\be
({\rm geodesic~length~in~dS\text{-}hemisphere}) = \pi \ell ~.
\ee
Thus, the corresponding Euclidean time separation $\Delta \tau_{dS}$ is always
\be
\Delta \tau_{dS} = \frac{\beta}{2} ~ .
\ee

\subsubsection*{Geodesic in centaur geometry}
\begin{figure}[t]
\begin{center}
\includegraphics[width=10.cm]{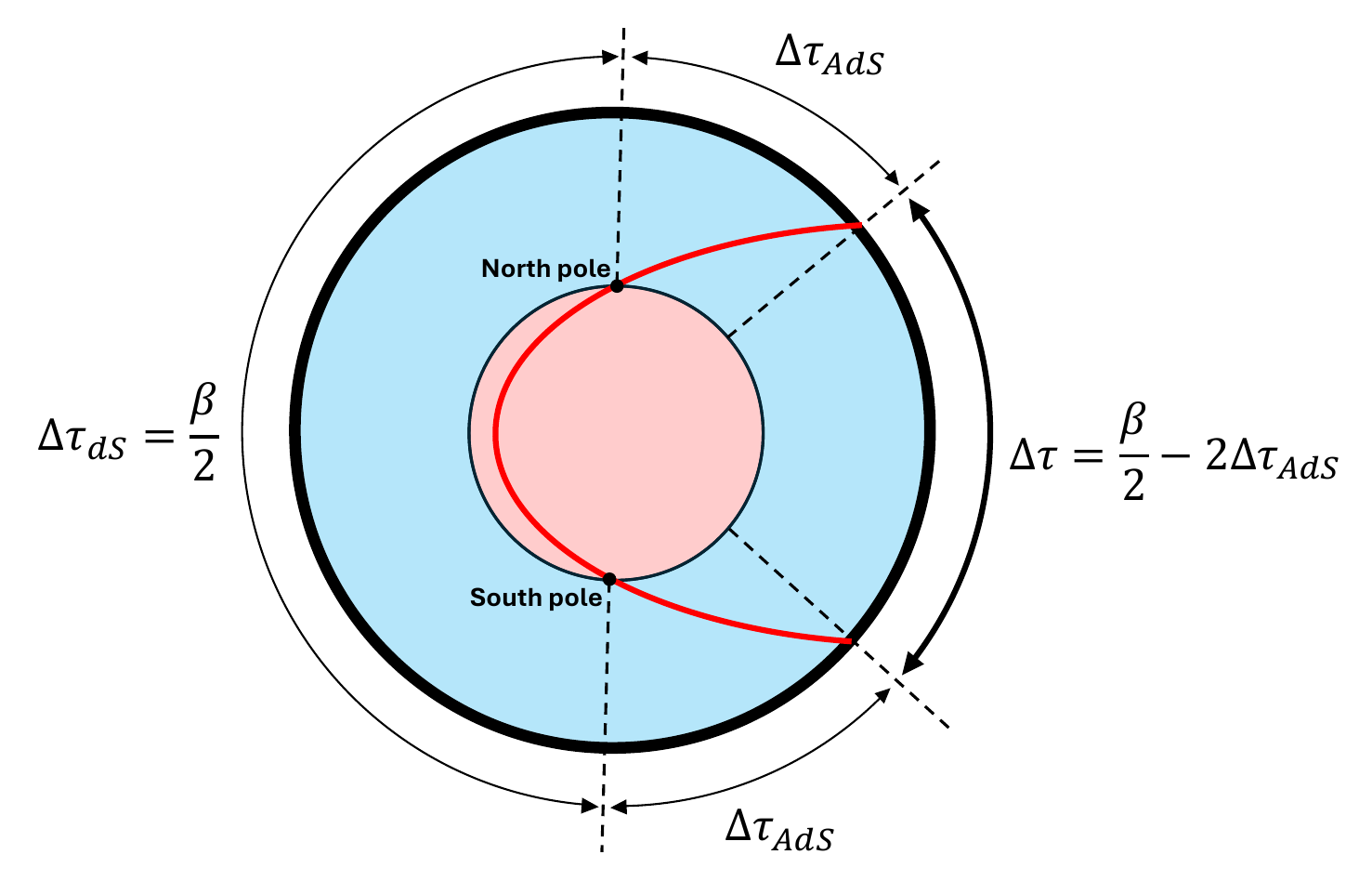}
\caption{Geodesic passing through the dS region with $\Delta \tau_{AdS}<\beta/4$. In this case, $\Delta \tau$ is given by $\Delta \tau = \frac{\beta}{2} - 2\Delta \tau_{AdS}$.}
\label{fig:geodesicCTR}
\end{center}
\end{figure}

Let us combine the two segments of geodesics in the AdS region and the single segment in the dS region. First, we focus on the case of small $\Delta \tau_{AdS}$, namely $\Delta \tau_{AdS} < \frac{\beta}{4}$. Due to our definitions of $\Delta \tau_{AdS}$ and $\Delta \tau_{dS}$, the relation among the boundary separation $\Delta \tau$, $\Delta \tau_{AdS}$, and $\Delta \tau_{dS} = \beta/2$ is given by (Fig.~\ref{fig:geodesicCTR})
\be
\Delta \tau = \frac{\beta}{2} - 2\Delta \tau_{AdS} ~ , \qquad (\Delta \tau_{AdS} < \frac{\beta}{4})~.
\label{eq:deltau0}
\ee

\begin{figure}[t]
\begin{center}
\includegraphics[width=8.cm]{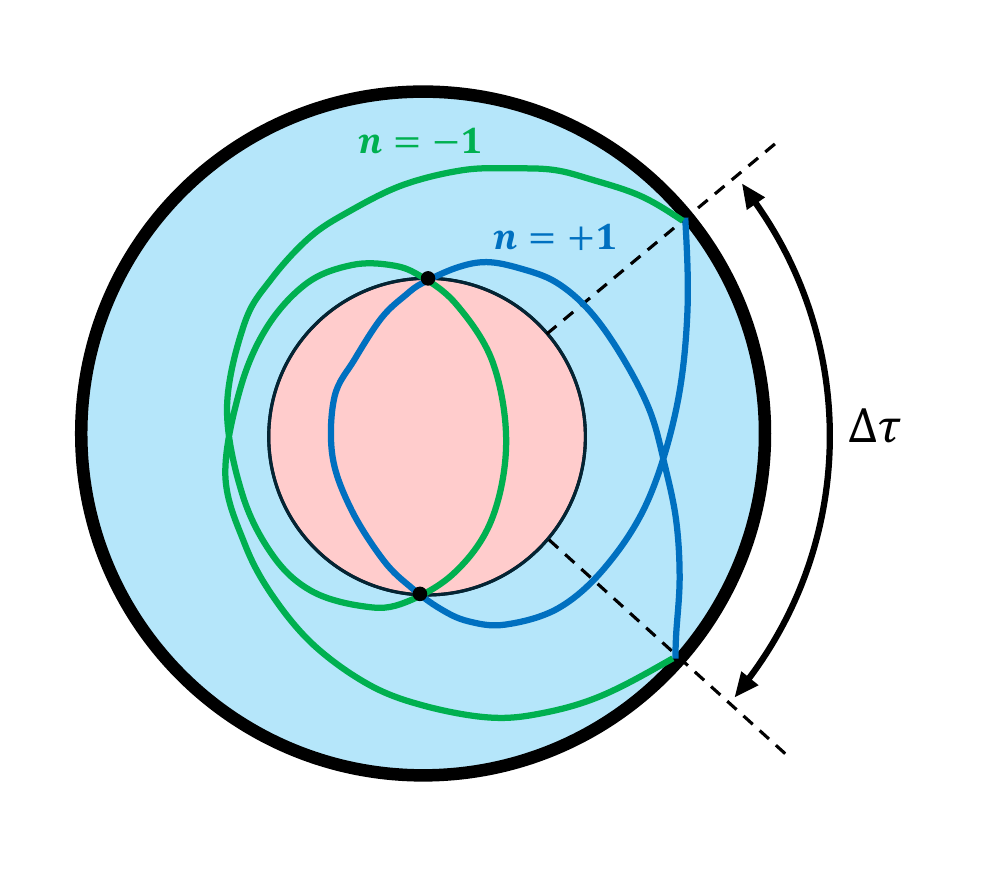}
\caption{Geodesics passing through the dS region with integer parameter $n=+1$ (blue) and $n=-1$ (green).}
\label{fig:geodesicCTR2}
\end{center}
\end{figure}

Similar to the previous case in which geodesics do not enter the dS region, there are infinitely many geodesics with the same boundary separation $\Delta \tau$. In this case, $\Delta \tau_{AdS}$ in the functions \eqref{eq:rofs} and \eqref{eq:tauofs} is related to $\Delta \tau$ by
\be
\Delta \tau_{AdS} = \left| \frac{\Delta \tau}{2} - \frac{\beta}{4} + \frac{\beta}{2}n \right| ~ ,
\ee
with an integer parameter $n\in \mathbb{Z}$. The case $n=0$ corresponds to \eqref{eq:deltau0} and Fig.~\ref{fig:geodesicCTR}, while the cases $n=-1$ and $n=+1$ are depicted in Fig.~\ref{fig:geodesicCTR2}.

As we will see shortly in the next subsection and in \eqref{eq:OSactionDS}, the geodesic length in the AdS region is given by
\be
({\rm geodesic~length~in~AdS~region}) \propto \log \left( \cosh\left( \frac{2\pi \Delta \tau_{AdS}}{\beta} \right) \right).
\ee
Therefore, the $n=0$ geodesic is the shortest one.

\subsection{Two-point function} 
\label{subsec:twoptfunc}
In the large mass limit $m \gg 1/\ell$, two-point functions in the bulk, and hence when the two points are taken to the boundary, the boundary two-point functions are approximated by geodesics. Here, we consider the Euclidean two-point function with the boundary separation $\Delta \tau$:
\be
G_{\beta}(\Delta \tau) \simeq e^{-I^{\rm p}|_{\rm on-shell}} ~ , \label{eq:G}
\ee
where $I^{\rm p}[\gamma]$ is the particle action and $I^{\rm p}|_{\rm on-shell}$ is the one evaluated on the dominant on-shell trajectory. The particle action consists of two terms: one is the bare action $I_{\rm bare}^{\rm p}[\gamma]$ and the other is the counterterm $I_{\rm ct}^{\rm p}(\partial \gamma)$:
\begin{align}
I^{\rm p}[\gamma] = I_{\rm bare}^{\rm p}[\gamma] + I_{\rm ct}^{\rm p}(\partial \gamma) ~ , \hspace{2cm} \label{eq:totalaction}  \\
I_{\rm bare}^{\rm p}[\gamma] = m \int_{\gamma} \sqrt{h} ds ~ , \hspace{1.6cm} \label{eq:bareaction} \\
I_{\rm ct}^{\rm p}(\partial \gamma) = \left. - m \ell \log \left( \frac{2 \ell \Phi}{\mathcal{C}} \right) \right|_{\partial \gamma} ~ . \label{eq:counteraction}
\end{align}
The counterterm is necessary since the geodesic length whose endpoints anchor the AdS boundary is always divergent and requires some renormalization. The constant $\mathcal{C}$ is an arbitrary constant and any choice of it works and determines the choice of the renormalization scheme. 

Let us compute the on-shell actions for geodesics staying in the AdS region and those for geodesics entering the dS region. The former is computed as
\footnote{
Similar calculations in JT gravity will be performed in detail in Appendix \ref{App:JTtwopt}. It will be helpful to understand what kind of computation is done here.
}
\be
I^{\rm p}|_{\rm on-shell} = 
- 2m \ell \log \left( \frac{2\pi \ell^2 \widetilde{\Phi}_{b} }{\mathcal{C} \beta} \frac{1}{\sinh( \frac{\pi \Delta \tau}{\beta})} \right) ~ , ~~~~ ({\rm staying ~ AdS~,}) \label{eq:OSactionADS}
\ee
and the latter is
\be
I^{\rm p}|_{\rm on-shell} = 
 - 2m \ell \log \left( \frac{2\pi \ell^2 \widetilde{\Phi}_{b} }{\mathcal{C} \beta} \frac{1}{\cosh( \frac{2 \pi \Delta \tau_{AdS}}{\beta})} \right) + m \pi \ell ~ . ~~~~ ({\rm entering ~ dS~.}) \label{eq:OSactionDS}
\ee
These confirm the previous statement that among the infinitely many geodesics, the $n=0$ ones are the shortest in each class. Therefore, $\Delta \tau_{AdS}$ in \eqref{eq:OSactionDS} is written as $\Delta \tau_{AdS} = \frac{\beta}{4} -\frac{\Delta \tau}{2} $, and thus 
\be
I^{\rm p}|_{\rm on-shell} = 
 - 2m \ell \log \left( \frac{2\pi \ell^2 \widetilde{\Phi}_{b} }{\mathcal{C} \beta} \frac{1}{\cosh\left( \frac{\pi}{2} - \frac{\pi \Delta \tau}{\beta} \right)} \right) + m \pi \ell ~ . ~~~~ ({\rm entering ~ dS}) \label{eq:OSactionDS2}
\ee

\begin{figure}[t]
\begin{center}
\includegraphics[width=5.cm]{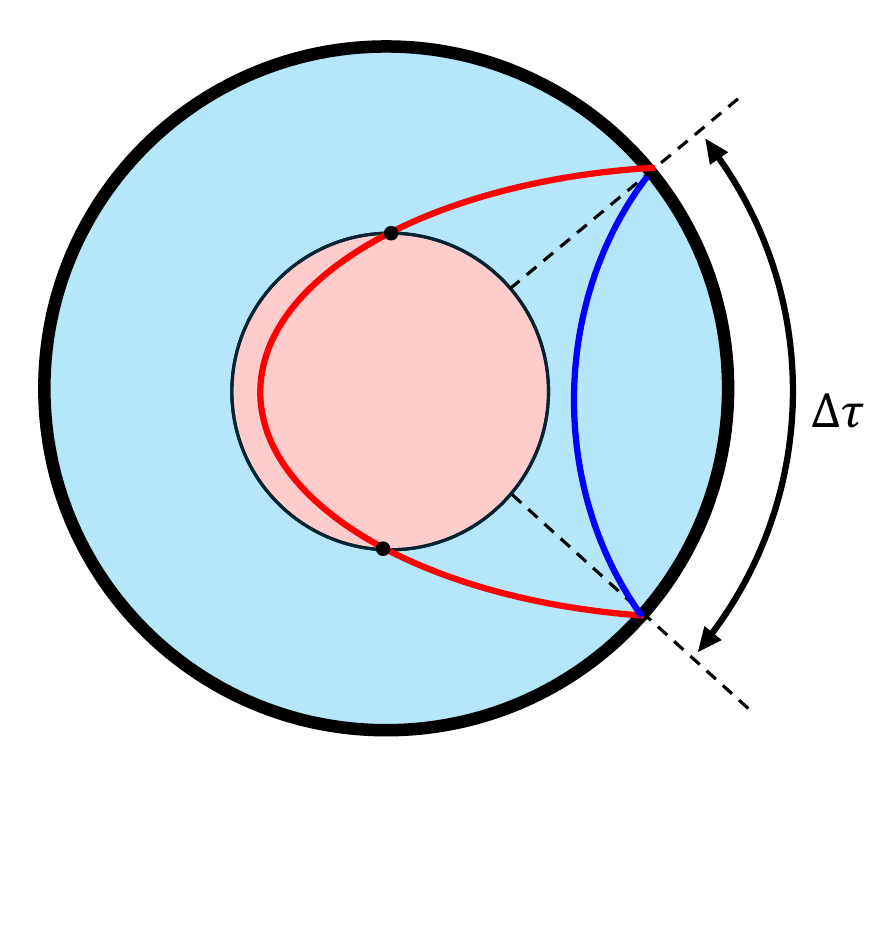} \includegraphics[width=10.cm]{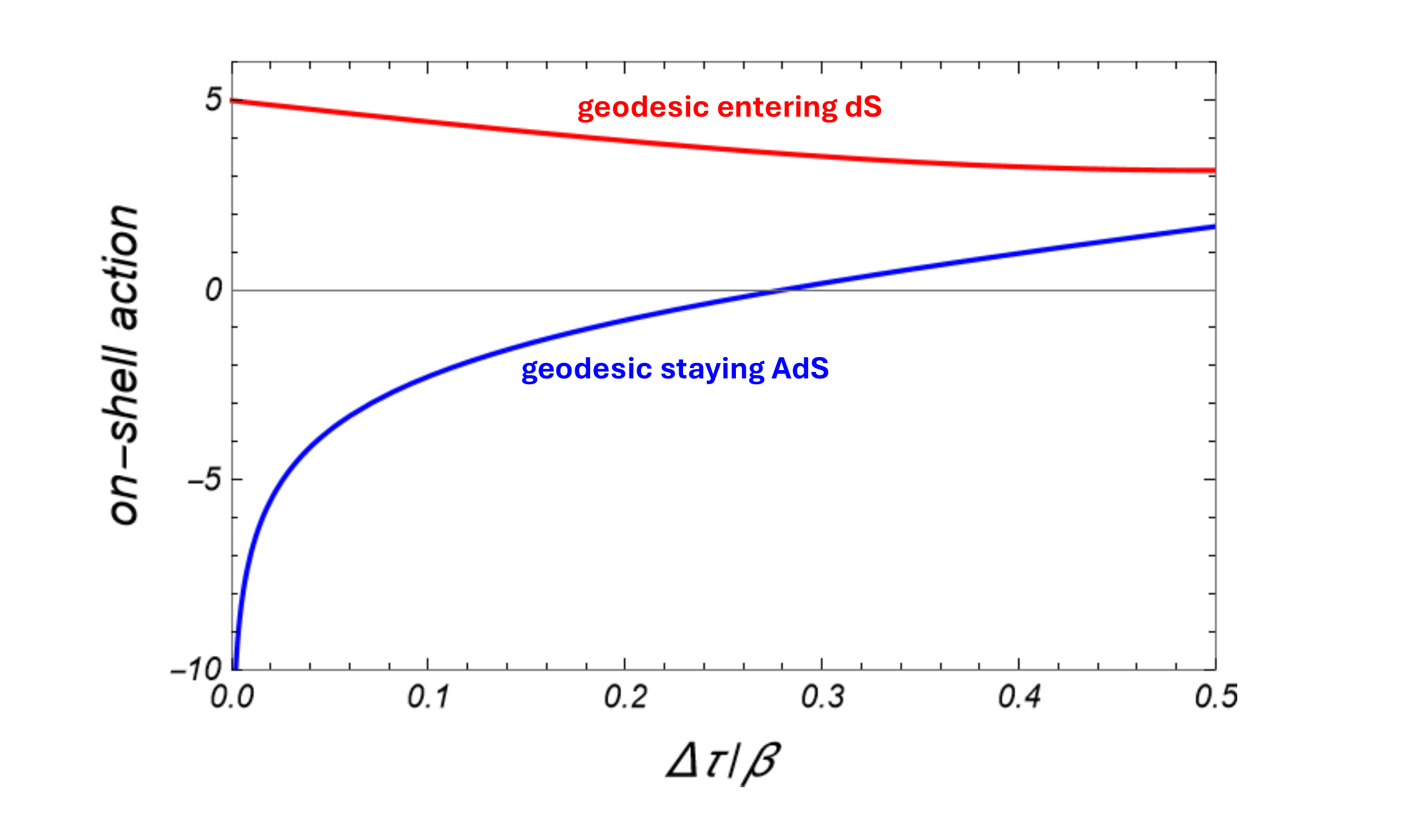} 
	\caption{(\textbf{Left}) Two geodesics with common boundary separation $\Delta \tau$, one staying in the AdS region (blue) and the other entering the dS region (red).  (\textbf{Right}) $\Delta \tau$ dependence of the on-shell actions \eqref{eq:OSactionADS} and \eqref{eq:OSactionDS2}. We set $m \ell=1$ and subtract the common factor $-2 m \ell \log \left( \frac{2\pi \ell^2 \widetilde{\Phi}_{b}}{\mathcal{C} \beta} \right)$. }
\label{fig:twogeodesic}
\end{center}
\end{figure}

So the remaining question is, for a given $\Delta \tau$, which geodesic is the shortest. (Fig. \ref{fig:twogeodesic} (\textbf{Left})). The comparison is shown in Fig. \ref{fig:twogeodesic} (\textbf{Right}). As can be seen from the figure, the geodesic staying in the AdS region is always shorter. Therefore, the one entering the dS region will be subdominant and the two-point function \eqref{eq:G} is given by
\be
\label{centaur2pt}
G_{\beta}(\Delta \tau) \simeq \left( \frac{2\pi \ell^2 \widetilde{\Phi}_{b} }{\mathcal{C} \beta} \frac{1}{\sinh( \frac{\pi \Delta \tau}{\beta})} \right)^{2 m \ell} ~.
\ee

\section{Information exchange}
\label{sec:5}
Having determined the Euclidean two-point functions in the centaur geometry, we now evaluate the entropy-change formula that quantifies the information exchange carried by a Hawking pair. Sec. \ref{sec:difference} briefly presents the necessary algebraic elements involving dressed operators. In Sec. \ref{sec:Sgrav}, we compute the gravitational contribution $\Delta S_{\rm grav}$ via the replica trick. Finally, Sec. \ref{sec:araki} evaluates the field-theory relative entropy $\Delta S_{\rm rel}$ using the Araki formula. Combining both pieces yields an inverse mini-Page curve for $\Delta S$ that peaks at $\tau \approx \beta/8$, providing the precise Euclidean-time scale at which information escapes the cosmological horizon purely from the algebraic structure of the crossed-product algebra.

\subsection{The entropy difference $\Delta S$}
\label{sec:difference}
Recall from Sec.~\ref{sec:centaur} that the centaur-algebra $\mathcal{A}$ has been promoted to the Type II$_\infty  $ von Neumann algebra $\hat{\mathcal{A}}$ via the crossed-product construction. The probe state is the Hawking-pair state
\be
\label{Hstate}
\ket \Phi = N^{-1/2} \phi_\tau \bar{\phi}_\tau \ket \Psi ~,
\ee
where $\ket \Psi$ is the thermofield-double reference state. The key algebraic result is the entropy difference
\be
\Delta S = \Delta S_{\rm grav}(\Phi|\Psi
) - \Delta S_{\rm rel}(\Phi||\Psi)~,
\ee
which provides a precise measure of the quantum information extracted by the pair and cleanly separates the gravitational and field-theory contributions.

In the crossed-product algebra $\hat{\mathcal{A}}$, the Hawking-pair operators $\phi_\tau$ and $\bar{\phi}_\tau$  are gauge-invariant. Explicitly, the dressed right-moving operator is
\be
\phi_\tau = e^{i p h_\Psi} \phi_\tau^{0} e^{-i p h_\Psi}~,
\ee
where $\phi_\tau^{0}$ belongs to the field theory algebra and commutes with the clock variables $q$ and $p$.  Let the unitary operator be $U = e^{i p h_\Psi}$. Since $U h_L U^\dagger = q$ with $h_L = q - h_\Psi$, we have
\be
h_L \phi_\tau = U q \phi_\tau^{0} U^\dagger = U \phi_\tau^{0} q U^\dagger = \phi_\tau h_L~,
\ee
and therefore $[h_L, \phi_\tau] = 0$. Similarly, for the mirror Hawking partner $\bar{\phi}_\tau$ we have $[h_R,\bar{\phi}_{\tau}]= 0$, which is consistent with the fact that mirror operators evolve in the opposite direction under modular flow generated by $h_\Psi$ ($\tilde{\tau} = \beta/2 - \tau$). 

With the dressed operators at hand, the four-point function that determines the normalization $N_{\beta,\tau} = \langle \Psi | \phi_{-\tau} \bar{\phi}_{-\tau} \phi_{\tau} \bar{\phi}_{\tau} | \Psi \rangle$ and both $\Delta S_{\rm grav}$ and $\Delta S_{\rm rel}$ is evaluated in free-field theory by Wick contractions. Because the centaur correlator $G_\beta(\Delta\tau)$ already incorporates the physics of the dS-interior and infrared effects, all subsequent contractions automatically carry the centaur geometry’s imprint. We proceed to evaluate the entropy change $\Delta S$ in the next subsections.

\subsection{Gravity contribution}
\label{sec:Sgrav}
We now compute the gravity contribution to the entropy change $\Delta S$  between the reference TFD state $\ket{\hat{\Psi}}$ and the probe Hawking-pair state $\ket{\hat{\Phi}}$ (see Sec. \ref{sec:crossed}). In the crossed-product Type II$_\infty$ algebra $\hat{\mathcal{A}}$ this contribution is defined as the expectation value of the right Hamiltonian
\be
\Delta S_{\rm grav} (\Phi, \Psi)= \bra{\hat{\Phi}} \beta h_R \ket{\hat{\Phi}}~,
\ee
where $h_R = (h_\Psi + X)/2$. 

The most direct route to compute it is the replica trick\footnote{Strictly speaking, we only replicate the inverse temperature $\beta$ not the operator insertions.}, which isolates the expectation value of the modular Hamiltonian by differentiating with respect to an auxilliary inverse-temperature parameter
\be
\Delta S_{\rm grav} (\Phi, \Psi) = -n \partial_n \bra{\hat{\Phi}} e^{\beta (1-n) h_R}\ket{\hat{\Phi}} \Big\lvert_{n=1}
= -n \partial_n \log \Big(  {\rm{Tr~}} \rho_{\h\Phi}~ e^{\beta(1-n)h_R} \Big) \Big\lvert_{n=1}~.
\ee
The replica trick is applied only to the inverse-temperature parameter $\beta$ appearing in the trace definition on the crossed-product Type II$_\infty$ algebra \label{eq:trace}. The overall minus sign is a conventional choice that yields the positive contribution expected for a black hole horizon in the canonical ensemble. 

\begin{figure}[t]
\begin{center}
\includegraphics[width=10.cm]{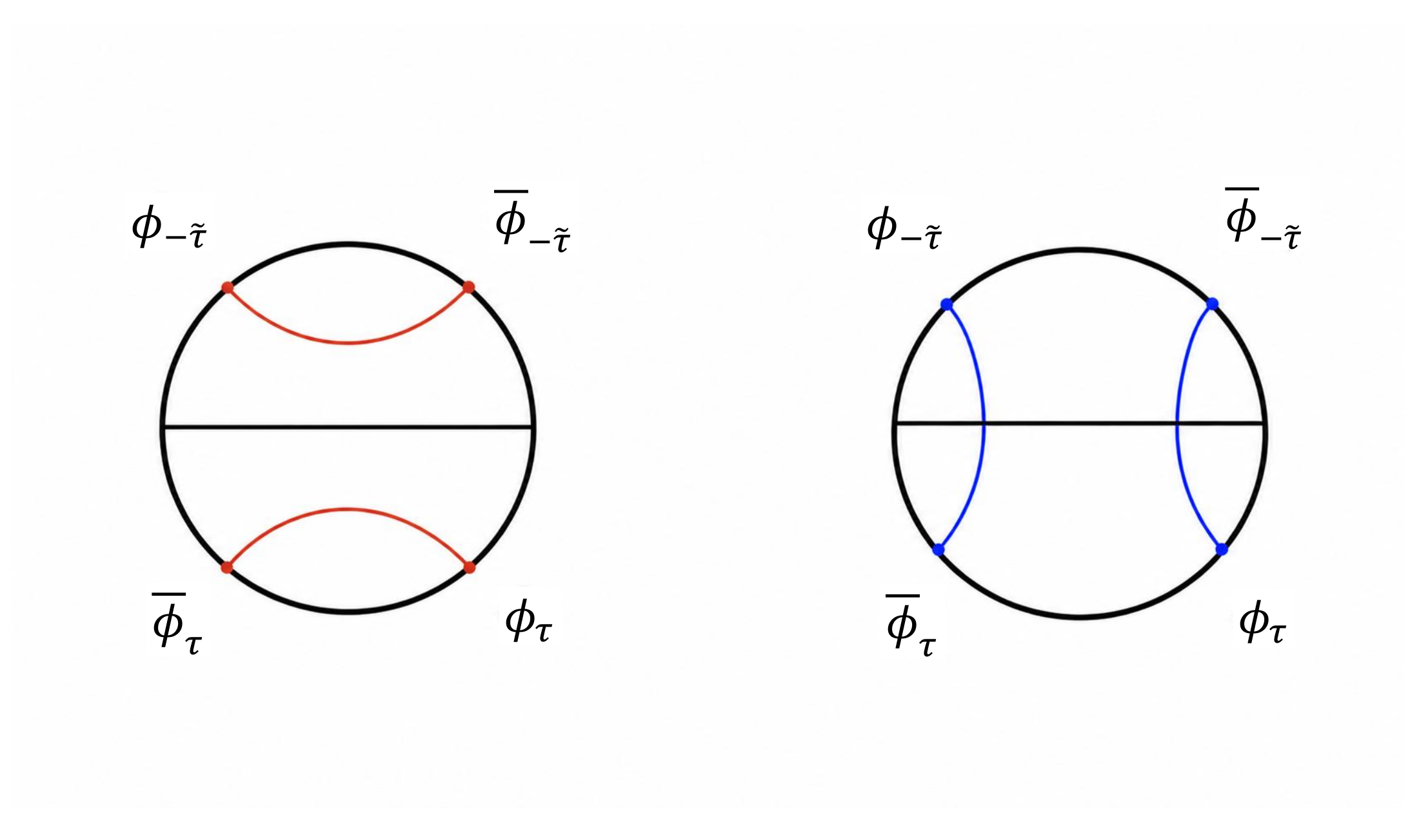} 
	\caption{Wick contractions.}
\label{fig:wick}
\end{center}
\end{figure}

To evaluate the trace explicitly we insert the state $|\hat{\Phi}\rangle$ and work in the free-field approximation. Using the commutators $[h_L,\phi_\tau]=0$ and $[h_R,\bar{\phi}_{\tilde{\tau}}]=0$, together with the KMS condition of the TFD state $|\Psi\rangle$ applied twice and the mirror relation $\bar{\phi}_{\tau}=J_\Psi\phi_\tau J_\Psi$, that identifies the interior partner with the reversed-flow exterior operator, we obtain the specific Wick contractions
\be\label{eq:wick}
\wick{\c1{\,{\phi}}_{\tau}\, \c1{{\bar\phi}_{\spc\tau}}} = 
G_{n\beta}(\tilde{\tau} - \tau)~, ~~ \wick{\c1{{\bar\phi}}{}_{-\tau} \spc\!\c1{\,
{\phi}}{}_{\tau}}= G_{n\beta}(2\tau)~.
\ee
These Wick contractions are depicted in Fig. \ref{fig:wick}.

The dominant pairing is the one whose time separation corresponds to the shortest bulk geodesic in the centaur geometry. In the free-field approximation the thermal four-point function therefore factorizes into products of thermal two-point functions evaluated at replica temperature $n\beta$ on this geodesic. This yields
\be
\operatorname{Tr}\bigl(\rho_{\hat{\Phi}}~\,e^{\beta(1-n)h_R}\bigr)=\frac{G_{n\beta}(y)^2}{G_\beta(y)^2}~,
\label{eq:wickfinal}
\ee
where the coordinate $y=\min(2\tau,\tilde{\tau} - \tau)  $ selects the dominant Wick pairing.

Substituting back into the replica formula and performing the derivative with respect to $n$ using the explicit centaur propagator from Sec.~\ref{sec:twopoint} gives
\be\label{eq:Sgrav}
\Delta S_{\rm grav}~ = -n\partial_n\log\bigl[G_{n\beta}(y)^2\bigr]\Big|_{n=1} = 4 m \ell \left( 1 - \frac{\pi y}{\beta} \coth\left( \frac{\pi y}{\beta} \right) \right)~.
\ee

To understand the physical content of Eq.~\eqref{eq:Sgrav} we examine its limiting behavior as a function of the Euclidean time $\tau$ of the Hawking pair. In the near-horizon regime ($\tau \to \frac{\beta}{4}$) we have $(\pi y/\beta)\coth(\pi y/\beta)\to 1$, hence $\Delta S_{\rm grav}\to0$: the pair is created essentially on top of the cosmological horizon, the entanglement remains maximal, and no information has yet come out from the dS interior. At the transition point ($\tau =\beta/8$) the function $1-(\pi y/\beta)\coth(\pi y/\beta)$  attains its extremum; this marks the precise moment of maximum information transfer. Finally, in the far-from-horizon regime ($\tau \to 0$) we find $(\pi y/\beta)\coth(\pi y/\beta)\to 1$, so $\Delta S_{\rm grav}\to 0$ again. In this case, the Hawking partner has propagated far into the exterior and the mirror operator is now fully disentangled from the interior degrees of freedom. The resulting inverse Page-like curve therefore encodes the complete microscopic timeline on which information escapes the cosmological horizon in the centaur geometry.

\begin{figure}[t]
\begin{center}
\includegraphics[width=10.cm]{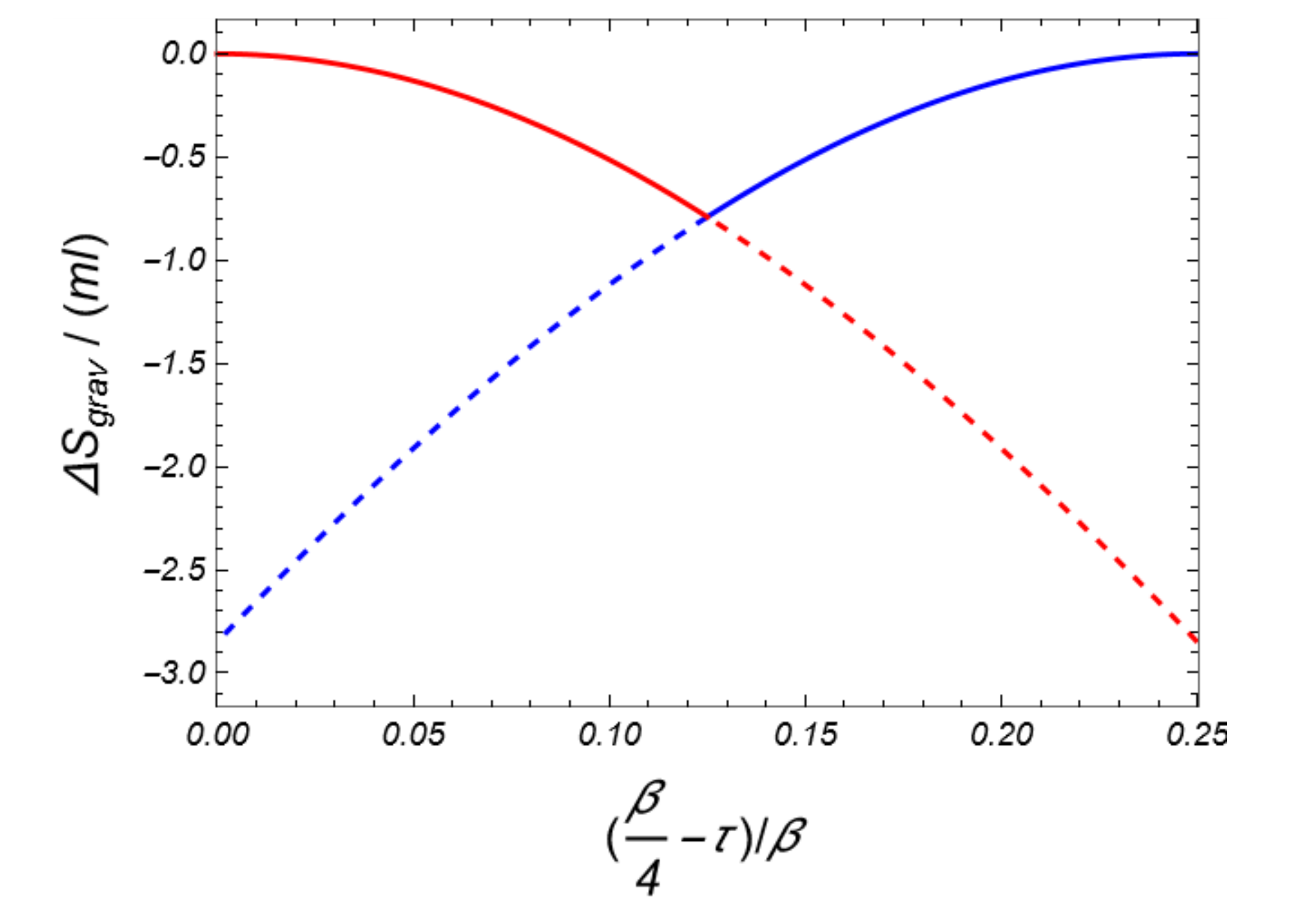} 
	\caption{``Inverse" mini-Page curve in the canonical ensemble. The near-horizon behavior, corresponding to $\tau$ close to $\beta/4$, is shown by the thick red curve. The far-away behavior, corresponding to $\tau$ close to $0$, is shown by the thick blue curve. The ``mini-Page time" is $\tau=\beta/8$.}
\label{fig:minipage}
\end{center}
\end{figure}

The resulting curve (Fig.~\ref{fig:minipage}) is the \emph{inverse} mini-Page curve for cosmological horizons in flow geometries. In contrast to the usual black-hole Page curve (which rises as information scrambles out and then falls once the radiation purifies), the hyperbolic propagator of the centaur geometry produces the inverse behaviour. Physically this reflects the distinctive thermodynamics of a cosmological horizon: the static patch has negative specific heat, so any added matter fluctuation decreases the total entropy. The algebraic construction shows that information transfer is not instantaneous at the horizon but occurs at a macroscopic Euclidean time $\tau\sim\beta/8$--precisely when the Hawking pair becomes real and disentangled from its mirror partner inside the cosmological horizon. This directly addresses the central open question of quantum information in dS spacetime: when does information escape the cosmological horizon? The mini-Page curve computed from the centaur crossed-product algebra supplies the precise microscopic timescale without an external bath or replica wormholes.

In the microcanonical ensemble the natural definition of the gravitational entropy difference is the expectation value of the shifted right-Hamiltonian 
\be
\Delta S_{\rm grav}(\Phi,\Psi) = \langle\hat{\Phi}|\beta(h_R-E)|\hat{\Phi}\rangle_{\rm micro}~.
\ee
Physically, the energy shift subtracts the fixed background energy of the reference TFD state. This isolates the extra energy carried away by the Hawking pair, which is precisely the gravitational contribution to the information content we wish to extract. 

Applying the replica trick to the microcanonical weight introduces an overall phase in the replicated partition function. 
\be
Z_n^{\rm micro} = \operatorname{Tr}\bigl(\rho_{\hat{\Phi}}^{\rm micro}\,e^{\beta(1-n)(h_R-E)}\bigr) \approx e^{-\beta(1-n)E} G_{n\beta}(y)^2~, 
\ee
where the second factor is the canonical replicated weight, which equals the insertion factor $G_{n\beta}(y)^2$ from free-field Wick contractions \eqref{eq:wickfinal}. Differentiating with respect to the replica index $n$ isolates the desired expectation value
\be
\Delta S_{\rm grav}(\Phi,\Psi) = -n\partial_n\log Z_n^{\rm micro}(\Phi)\Big|_{n=1} =  -n\beta E - n\partial_n\log G_{n\beta}(y)^2\Big|_{n=1}~.
\ee

\begin{figure}[t]
\begin{center}
\includegraphics[width=10.cm]{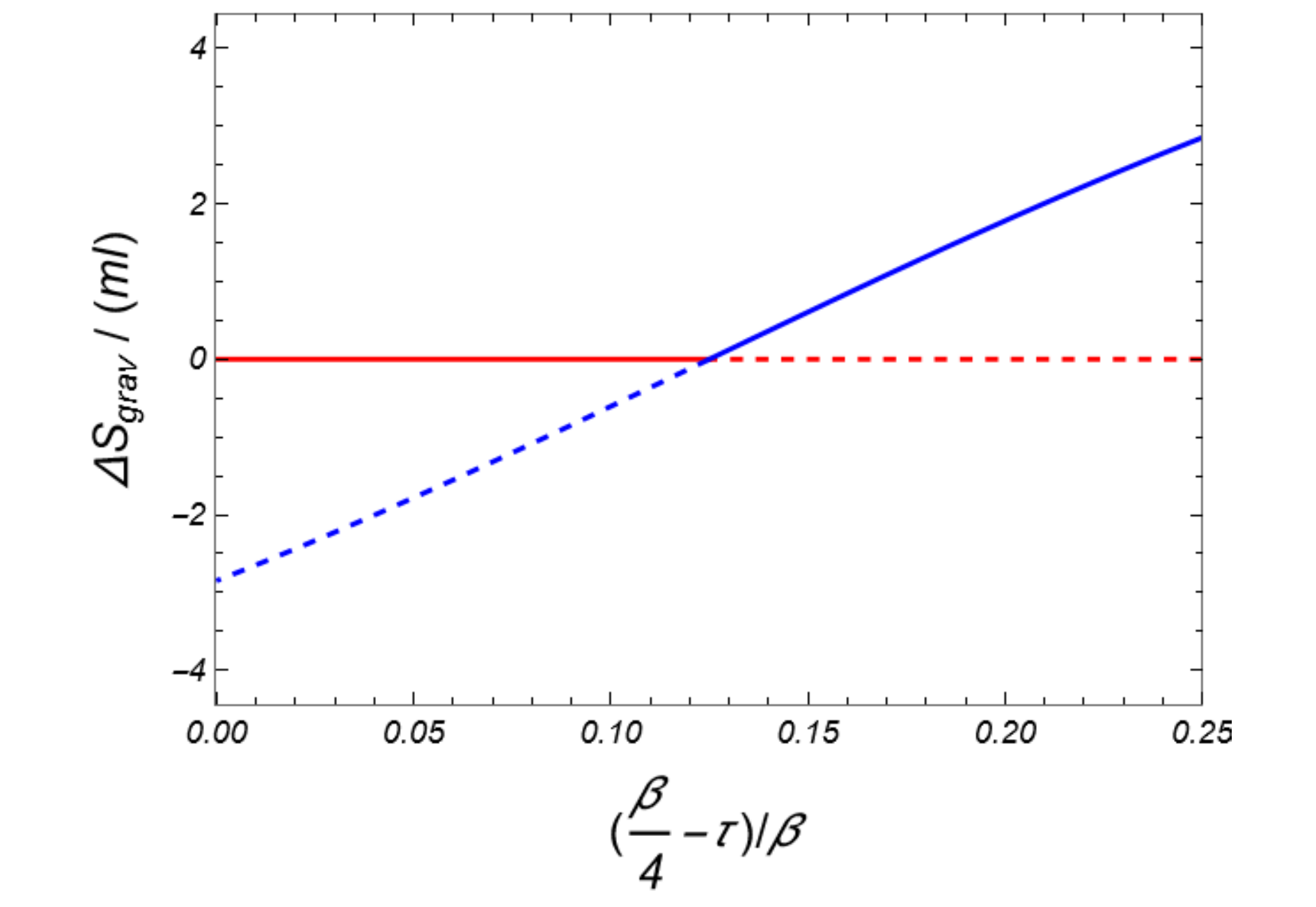} 
	\caption{ $\tau$ dependence of $S_{\rm grav}$ in microcanonical ensemble.}
\label{fig:MCcurve}
\end{center}
\end{figure}

In order to obtain a measure of information transfer it is natural to evaluate the same probe density matrix $\rho_{\Phi}$ in two different regimes of the Euclidean-time interval $0\leq\tau\leq\beta/4$. Split the interval into two regions separated by the transition point $\tau_c=\beta/8$ (midway between the horizon location $\tau=\beta/4$ and the asymptotic AdS boundary at $\tau=0$), evaluate the replica derivative separately in the near horizon $\beta/4>\tau>\beta/8$ and far-away regions respectively, and subtracting, we find
\begin{align}
\Delta S_{\rm grav}  =  \beta \partial_\beta \log \left( \frac{G_\beta(\tilde{\tau} -\tau )^2}{G_\beta(2\tau)^2} \right) \notag \hspace{6.2cm}  \\
 ~  =  \frac{2\pi h}{\beta}\left[ 2 \left( \frac{\beta}{2} - 2\tau \right) \coth(\frac{\pi \left( \frac{\beta}{2} - 2\tau \right)}{\beta}) -4\tau \coth(\frac{2\pi \tau}{\beta}) \right] \label{eq:microDeltaS}
\end{align}
The behavior of \eqref{eq:microDeltaS} that follows from the centaur propagator is shown in Fig. \ref{fig:MCcurve}, which shows a local increase in entropy when the Hawking pair becomes real. After the transition at $\tau_c = \beta/8$, $\Delta S$ increases because the dominant Wick contractions on the replica manifold shift from the mirror to the crossing channel (Fig. \ref{fig:wick}). Once the Euclidean separation exceeds this critical value, further separation no longer produces significant additional backreaction on the cosmological horizon; instead, the excess energy of the pair is transported outward along the future horizon and is accounted for by the growth of the entanglement wedge that tracks the emitted particle. This local rise in the probe entropy difference does not contradict the fact that positive-energy matter always reduces the size of the cosmological horizon in de Sitter space. The quantity $\Delta S$ isolates only the local excess energy carried by a single Hawking pair after subtraction of the fixed background energy; the total generalized entropy of the cosmological horizon remains bounded by the Gibbons-Hawking value. This boundedness is consistent with the unitary Page curve derived in global de Sitter holography, where the entropy of a subregion grows at most up to the horizon entropy before saturating \cite{Geng:2021wcq}.

\subsection{Araki formula}
\label{sec:araki}
Apart from the gravitational contribution $\Delta S_{\rm grav}$ derived in Sec. \ref{sec:Sgrav}, the entropy difference $\Delta S$ also contains the field-theory relative entropy term
\be\label{eq:Srel}
\Delta S_{\rm rel}(\Phi \lvert \lvert \Psi) =  \bra{\Phi} h_{\Psi \lvert \Phi} \ket{\Phi}~,
\ee
where $h_{\Psi|\Phi} = \log\Delta_{\Psi|\Phi}$ is the relative modular Hamiltonian. The Araki relative entropy between any two faithful normal states on a von Neumann algebra is well-defined for all types of algebras including Type III$_1$ and the Type II$_\infty$ algebra $\hat{\mathcal{A}}$. In our case it refers to the field-theory operator algebra $\mathcal{A}$. 

\begin{figure}[t]
\begin{center}
\includegraphics[width=12.cm]{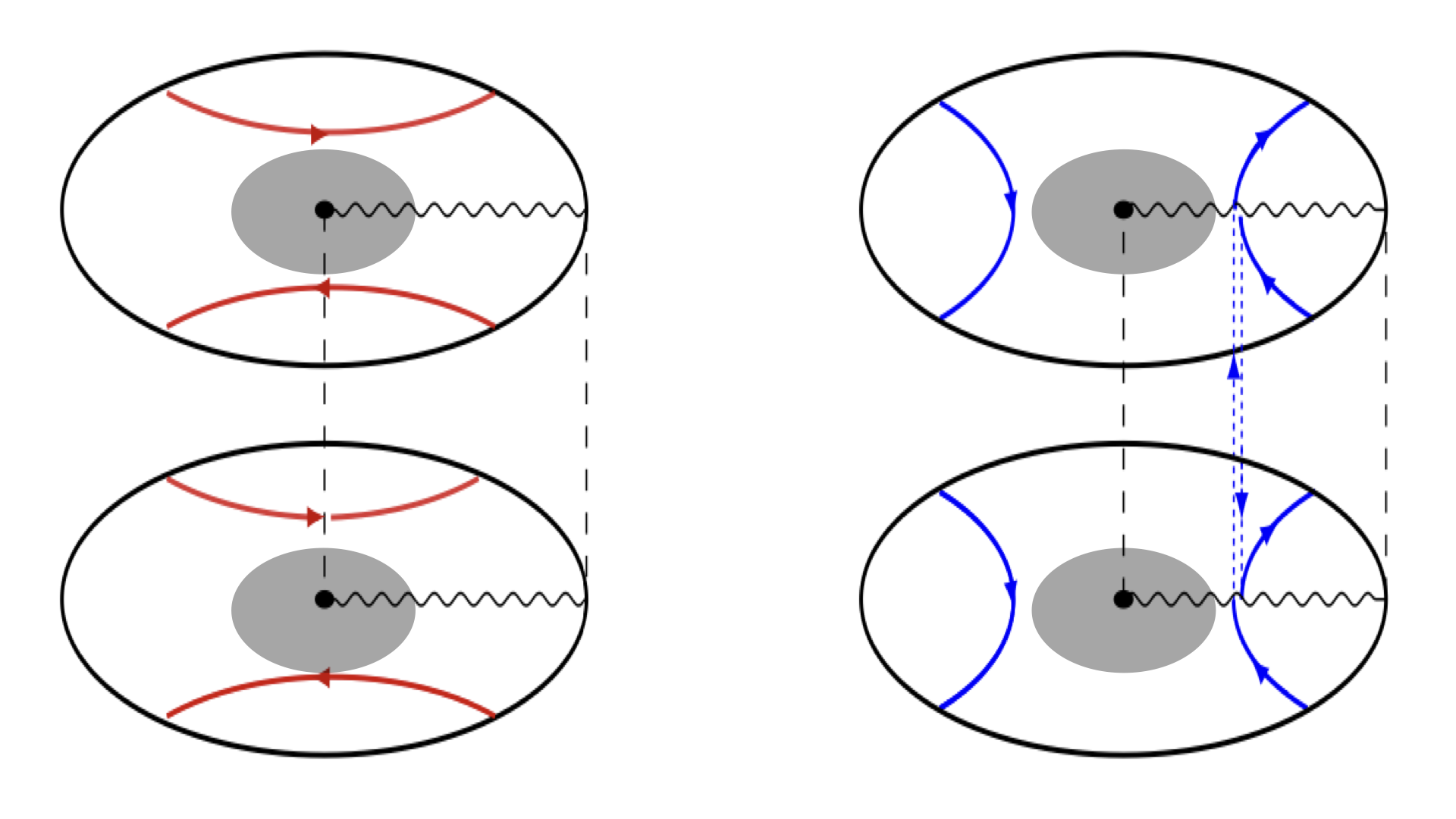} 
	\caption{Euclidean $n=2$ sheeted manifold.}
\label{fig:replica manifold}
\end{center}
\end{figure}

We compute $\Delta S_{\rm rel}$ using the field theory replica trick. The $n  $-th relative Rényi entropy is
\be\label{eq:Sreln}
\Delta S_{\rm rel,n}(\Phi \lvert \lvert \Psi) = \frac{1}{1-n}\log\langle\Phi|\Delta_{\Psi|\Phi}^{1-n}|\Phi\rangle~.
\ee
The Araki relative entropy $\Delta S_{\rm rel}(\Phi||\Psi)$ is recovered in the limit $n\to 1$. The Euclidean manifold $\Sigma_n$ that realizes this replica trick contains both the AdS-trumpet and dS-hemisphere regions of the centaur geometry (see Fig. \ref{fig:replica manifold}). A related Euclidean method for the centaur geometry was used in the computation of the flow Hilbert space in \cite{Espindola:2025wjf}. 

Let us first compute the $n=2$ case in full detail. We evaluate the matrix element $\bra\Phi\Delta_{\Psi|\Phi}^{-1}\ket\Phi$. Applying the cocycle identity once gives
\be
\bra\Phi\Delta_{\Psi|\Phi}^{-1}\ket \Phi = \bra\Phi\Delta_\Phi^{-1} \Delta_{\Phi|\Psi} \Delta_\Psi^{-1} \ket\Phi =\bra\Phi \Delta_{\Phi|\Psi} e^{\beta h_\Psi}\ket\Phi~,
\ee
where the second equality follows from $h_\Phi|\Phi\rangle=0  $ (\ie, $ \Delta_\Phi|\Phi\rangle=|\Phi\rangle  $) and the self-adjointness property of $\Delta_\Phi$. Inserting the definition of the Hawking pair state \eqref{Hstate} and using the identity for the relative Tomita operator $\langle\Psi|\b\Delta_{\Phi|\Psi}\a|\Psi\rangle = \langle\Phi|\a \b|\Phi\rangle$, which is valid for any operators $\a,\b\in\mathcal{A}$, we find
\be
\bra{\Psi}\phi_{-\tilde{\tau}}\bar{\phi}_{-\tau} \Delta_{\Phi|\Psi} e^{\beta h_\Psi} \phi_\tau\bar{\phi}_{\tilde{\tau}}\ket{\Psi} = \bra{\Phi}\phi_{\tau-\beta}\bar{\phi}_{\tilde{\tau}-\beta}\phi_{-\tilde{\tau}}\bar{\phi}_{-\tau}\ket{\Phi}~,
\ee
where the modular operator generates Euclidean time evolution $\beta$ along the thermal circle. Reordering the eight operators on the thermal circle via the Tomita mirror relation $\tilde{\tau} = \beta/2 - \tau$  and the periodicity of the Euclidean circle of size $\beta$ replaces each mirror time with its exterior counterpart shifted by $\beta/2$ 
\be 
\bra{\Psi}\phi_{\tau-\beta/2}\bar{\phi}_{-\tau}\phi_{\tau-\beta}\bar{\phi}_{\tilde{\tau}-\beta}\phi_{\tau+\beta/2}\bar{\phi}_{-\tau+\beta}\phi_\tau\bar{\phi}_{\tilde{\tau}}\ket{\Psi}~.
\ee
Applying the KMS condition once more cyclically reorders the operators into the nearest-neighbour Wick pairings, yielding
\begin{align}
\bra{\Phi}\Delta_{\Psi|\Phi}^{-1}\ket{\Phi}   & =\bra{\Psi}\phi_{\tau+\beta/2}\bar{\phi}_{-\tau}\phi_\tau\bar{\phi}_{\tilde{\tau}}\phi_{\tau-\beta/2}\bar{\phi}_{-\tau}\phi_{\tau-\beta}\bar{\phi}_{\tilde{\tau}-\beta}\ket{\Psi}~ \\ \nonumber
& = \bra{\Psi}\prod_{k=-1}^{2}\phi_{\tau-k\beta/2}\bar{\phi}_{\tilde{\tau}-k\beta/2}\ket{\Psi}~.
\end{align}

In order to extend this result to general $n$, we proceed inductively. Assuming the result holds for $n$ pairs, applying the cocycle identity to $\bra{\Phi}\Delta_{\Psi|\Phi}^{1-(n+1)}\ket\Phi$ and repeating similar steps as above (that is, the definition of the relative operator $\Delta_{\Phi|\Psi}$, the commutators $[h_L,\phi_\tau]=0$, $[h_R,\bar{\phi}_{\tau}]=0$, the KMS condition, and the mirror relation $\bar{\phi}_{\tau}=J_\Psi\phi_\tau J_\Psi$) inserts one additional pair $\phi_{\tau-n\beta/2}\bar{\phi}_{\tilde{\tau}-n\beta/2}$. Substituting the inductive hypothesis then produces the product of pairs in the state $\ket{\Psi}$ running from $k=-1, \dots ,2n$, completing the proof. In the replica-manifold interpretation we have
\be
\bra\Phi\Delta_{\Psi|\Phi}^{1-n}\ket\Phi = \frac{1}{Z_{\beta,\tau}^{2n-2}}\bra\Psi\prod_{k=-1}^{2n-2}\phi_{\tau-k\beta/2}\bar{\phi}_{\tilde{\tau}-k\beta/2}\ket\Psi~,
\ee
with $\Sigma_n$ the $n$-sheeted branched cover of the Euclidean centaur geometry and $Z_{\beta,\tau}$ the normalization factor of the Hawking-pair state. Importantly, in this formula the operator insertions are replicated but $\beta$ is not rescaled by a factor of $n$. The insertions span the full range of Euclidean times of an $n$-replica geometry (see Fig. \ref{fig:replica manifold})

We evaluate this $2n$-point correlator on $\Sigma_n$. In the free-field approximation the only non-zero contractions are the nearest-neighbour pairings \eqref{eq:wick}. Each term in the product is a single nearest-neighbour pair consisting of one exterior operator $\phi$ at Euclidean time $\tau - k\beta/2$ and its mirror partner $\bar{\phi}$ at $\tilde{\tau} - k\beta/2$. Consequently, every individual Wick contraction evaluates to
\be 
\langle\phi_{\tau-k\beta/2}\,\bar{\phi}_{\tilde{\tau}-k\beta/2}\rangle_{\Sigma_n}=G_\beta(y)~,
\ee
where $y=\min(2\tau,\tau-\tilde{\tau})$ and $G_\beta(y)$ is the centaur thermal two-point function \eqref{centaur2pt} derived from the AdS/dS geodesic analysis in Sec \ref{sec:twopoint}. The crucial point is that $G_\beta(y)$ does not depend on the replica index $n$, because the QFT replica trick does not rescale the temperature $\beta$.

Since all $2n$ pairings are identical and there are no other contractions (by the Wick theorem for free fields), the full correlator factors exactly as 
\be
\langle\Phi|\Delta_{\Psi|\Phi}^{1-n}|\Phi\rangle = G_\beta(y)^{2n} / Z_{\beta,\tau}^{n}~, \quad Z_{\beta,\tau}=\langle\Psi|\phi_{-\tilde{\tau}}\,\tilde{\phi}_{-\tau}\,\phi_\tau\,\tilde{\phi}_{\tilde{\tau}}|\Psi\rangle~.
\ee
In the free-field approximation $Z_{\beta,\tau}$ factorizes via Wick’s theorem into two channels, both evaluated with the dominant geodesic in the flow geometry: the mirror channel with separation $\tau-\tilde{\tau}$ contributing $  G_\beta(\tau - \tilde{\tau})$ and the crossing channel with separation $2\tau$ contributing $G_\beta(2\tau)$. 

In the near-horizon regime ($y\to 0$) the mirror pairing diverges as $G_\beta(y)\sim y^{-2m\ell}$, while the crossing pairing, $G_\beta(\beta/2)$, remains finite . Thus  $Z_{\beta,\tau}\approx G_\beta(y)^2$ and the powers cancel exactly giving $\langle\Phi|\Delta_{\Psi|\Phi}^{1-n}|\Phi\rangle\approx 1$. Similarly, the same exact cancellation holds in the far-away regime ($y\to\beta/2$). Here the crossing channel dominates, with $G_\beta(y)\sim(\beta/2-y)^{-2 m \ell}$, while the mirror channel remains finite. 

At the transition point $y=\beta/4$, corresponding to $\tau=\beta/8$, the mirror and the crossing channels become equal. Both contribute with identical strength, so $Z_{\beta,\tau}\approx  2\,G_\beta(y)^2.$ The replica matrix element is therefore
$\langle\Phi|\Delta_{\Psi|\Phi}^{1-n}\ket{\Phi} \approx2^{-n}.$ Taking the $n \to 1$ limit recovers the Araki relative entropy 
\be
\Delta S_{\rm rel}(\Phi||\Psi) \Big\lvert_{\tau \approx \beta/8}\approx\log2~.
\ee
This maximum entropy result represents the precise instant when the Hawking pair transfers its full information from the dS interior across the cosmological horizon in the centaur geometry. Physically, it encodes the peak uncertainty in the effective ``Hawking qubit'' describing the pair’s occupancy, where the operator-mirror entanglement is half-dissolved and the pair carries one bit of quantum information to the exterior observer. Away from this narrow transition region one Wick contraction channel dominates exponentially, driving the field-theory relative entropy rapidly to zero.

The analysis so far has been carried out in the canonical ensemble. To establish robustness of this result under a change of ensemble, we now consider the microcanonical setting in which the total ADM energy is fixed to a narrow window around a value $E$ ($\delta E\ll E$).

\paragraph{Inverse mini-Page curve} Consequently, to leading semi-classical order the field-theory contribution vanishes everywhere except in a narrow window around the transition point, and the full entropy difference $\Delta S$ between the reference TFD state and the state containing the Hawking pair is given exclusively by the gravitational contribution. The inverse mini-Page curve therefore encodes the complete microscopic timeline on which quantum information escapes the cosmological horizon into the exterior region, purely through the gravitational dressing of the Hawking pair within the crossed-product Type II$_\infty$ algebra. The analysis of the Araki formula thus demonstrates that the gravitational contribution $ \Delta S_{\rm grav}$ alone governs the information dynamics of Hawking pairs in cosmological horizons, providing a direct algebraic realization of when information is transferred across the cosmological horizon. 

\section{The generalized entropy}
\label{sec:Sgen}

\subsection{Algebraic entropy}
\label{sec:salgebraic}
So far we have worked entirely in the  canonical ensemble. We now switch to the microcanonical ensemble to analyze the physics of the Hawking-pair state $\ket{\Phi}$. We are interested in the microcanonical thermofield double state of the left and right boundary theories (see Fig. \ref{fig:centaur1}) given by
\be
\left|\Psi_{\rm TFD} \right\rangle = \sum_{a} f_{E}(E_{a}) |E_{a}\rangle_{L} | E_{a}\rangle_{R}~,
\ee
where $f_{E}(E_{a})$ is a window function sharply peaked around $E_{a}=E$. A generalization of this state into a partially entangled thermal state (PETS) \cite{Goel:2018ubv} was recently analyzed \cite{Espindola:2025wjf}, leading to a microscopic understanding of the Gibbons-Hawking entropy. In order to study semiclassical physics, we choose $f_{E}(E_{a})$ so that it does not have tails, that is, it has a sufficiently small compact support. This keeps energy fluctuations $  \mathcal{O}(1)$ even in the strict large-$N$ limit and allows the renormalized ADM Hamiltonian $ h_R = H_R - E_0$ to be adjoined directly to the algebra. The resulting algebra is the crossed product algebra
\be
\h \cA = \mathcal{A} \rtimes \mathbb{R}~,
\ee
generated by elements $\a~e^{i\alpha h_R}\otimes e^{i\alpha X}$ with $\a\in\mathcal{A}_R$ and $X$ the conjugate time-shift mode on $L^2(\mathbb{R})$. The extended Hilbert space is denoted by $\hat{\mathcal{H}}_{\rm Flow} = \mathcal{H}_{\rm Flow} \otimes L^2(\mathbb{R})$, where $\mathcal{H}_{\rm Flow}$ is the GNS representation of $\mathcal{A}$ on $  \ket\Psi$. We focus on semiclassical states with small time shift fluctuations, \ie, $\Delta p = O(\varepsilon)$,
\be
\ket{\hat{\Phi}} = \int_{-\infty}^\infty dx\,\sqrt{\varepsilon}\,g(\varepsilon x)\,|\Phi\rangle|x\rangle~,
\ee
where $g\in L^2(\mathbb{R})$ is normalized and the Hawking-pair state
\be
|\Phi\rangle = N^{-1/2}\phi_\tau\bar{\phi}_{\tilde{\tau}}|\Psi\rangle,\qquad N = G(v)
\ee
is labeled explicitly by the outgoing null coordinate $v$ on the future cosmological horizon (the two-point function $G(v)$ is the horizon correlator computed below). Importantly, the renormalized right Hamiltonian acts as $ \beta h_R = \beta(x + \hat{h})$ with $  \hat{h} = -\log\Delta_\Psi  $, and the trace on $  \hat{\mathcal{A}}  $ has the form
\be
\operatorname{Tr}[\hat{\a}] = \int_{-\infty}^\infty dx~e^{\beta x}\langle\Psi|\hat{\a}|\Psi\rangle~.
\ee
The candidate density matrix associated to the algebra $ \hat{\mathcal{A}}  $ can be derived following \cite{Chandrasekaran:2022cip}
\be
\rho_{\hat{\Phi}}\approx\varepsilon\,g(\varepsilon h_R)\,e^{-\beta x}\,\Delta_{\Phi|\Psi}\,g(\varepsilon h_R)\qquad(+ O(\varepsilon)\text{ corrections})~,
\ee
where $\Delta_{\Phi|\Psi} = S_{\Phi|\Psi}^\dagger S_{\Phi|\Psi}  $ is the relative modular operator on $  \mathcal{A}$. Importantly, the trace condition holds to $  O(\varepsilon)  $ by the standard Fourier analysis on generic element of $  \hat{\mathcal{A}}  $.

Since $g(\varepsilon h_R)$ commutes with $  \Delta_{\Phi|\Psi}  $ to leading order we have
\be
\log\rho_{\hat{\Phi}}\approx -\beta x - h_{\Phi|\Psi} + \log\bigl(\varepsilon|g(\varepsilon h_R)|^2\bigr) + O(\varepsilon)~.
\ee
The von Neumann entropy on $  \hat{\mathcal{A}}  $ therefore becomes
\be
S(\hat{\Phi}(v))_{\tilde{\mathcal{A}}_R} = \langle\hat{\Phi}|\beta h_R|\hat{\Phi}\rangle - \langle\hat{\Phi}|h_{\Phi|\Psi}|\hat{\Phi}\rangle - \langle\hat{\Phi}|\log\varepsilon|g(\varepsilon h_R)|^2|\hat{\Phi}\rangle + O(\varepsilon)~.
\ee
The first term is the gravity contribution $  S_{\rm grav}(\Phi|\Psi)$ due to the Hamiltonian of the right exterior; the second term is precisely the Araki relative entropy $  S_{\rm rel}(\Phi||\Psi)$. 

This entropy has a transparent local interpretation in terms of a single emitted Hawking particle. The algebra $\hat{\mathcal{A}}$ consists of all observables measurable from the right AdS$_2$ boundary (a Type II$  _\infty  $ factor). Its von Neumann entropy given by $S(\hat{\Phi}(v))_{\hat{\mathcal{A}}}$ therefore counts the quantum information that has been carried out by one Hawking particle emitted along the outgoing null ray, as seen by a right-exterior observer.

In order to relate the von Neumann entropy with the generalized entropy, we adapt Wall’s argument to two-dimensional dilaton gravity. At future infinity on the cosmological horizon the bulk state becomes indistinguishable from the background state $\ket\Psi  $, so $  S_{\rm bulk}(\infty)_\Phi = S_{\rm bulk}(b)_\Psi  $. Let $v$ be the affine parameter along the future-directed null generators with $v=0$ at the bifurcation surface. The null expansion $\theta = 2\partial_v\log\Phi  $ satisfies the two-dimensional Raychaudhuri equation
\be
\frac{d\theta}{dv} = -8\pi G\,T_{vv} + \frac{\kappa}{2}\theta - \frac12\theta^2~.
\ee
Because the dS$_2$ interior has $\kappa > 0  $, the dilaton contracts when the congruence is deformed inwards in the null direction ($  \theta < 0  $ for outgoing generators). Integrating from $  v=0  $ to $  v\to\infty  $ ($  \theta\to0  $) yields
\be
\frac{\Phi(\infty)}{4G} - \frac{\Phi(b)}{4G} = \beta\langle\Phi|\hat{h}_R|\Phi\rangle + O(G)~.
\ee
At future infinity $  \Phi(\infty)/4G = \Phi_0/4G + \beta h_R  $, so the relative-entropy term cancels the area difference and we obtain
\be
S_{\rm rel}(\Phi||\Psi) = S_{\rm gen}(\infty) - S_{\rm gen}(b)~.
\ee
Substituting back into the entropy formula immediately gives
\be
S(\hat{\Phi}(v))_{\hat{\mathcal{A}}} = S_{\rm gen}(b(v)) + {\rm const}~.
\ee
This equation identifies the algebraic entropy with the generalized entropy of the entanglement wedge cut $b(v)$ of the right-exterior algebra. In the centaur geometry $b(v)$ corresponds to the bifurcation surface of the cosmological horizon at the precise affine emission time $v$ of the particle. Its location moves outward along the horizon exactly tracking the emitted outgoing mode. The bulk term $S_{\rm bulk}(b)$ is encoded in the relative entropy $S_{\rm rel}  $, while the area term $\Phi(b)/4G$ follows from the gravitational dressing.

\subsection{Relative modular flow}
We now examine the real-time dynamics of the emitted Hawking pair on the cosmological horizon. The relative modular operator between the perturbed Hawking-pair state and the reference TFD state encodes the precise algebraic mechanism that governs how the pair disentangles from the dS interior. While the relative modular operator itself does not in general generate an automorphism group of the algebra $\cA$, an effective relative modular flow can be defined through its action on correlation functions in the perturbed state. We derive this effective flow from first principles and evaluate its action on horizon operators explicitly in the near horizon region.

In the near horizon regime with radial coordinate $\rho = r - r_h$ ($\rho \ll \ell$), the metric of the sharp centaur geometry (\ref{eq:dShemisphere}) reduces to the Rindler form
\be
ds^2 = -\Bigl(\frac{2\pi\tilde{\rho}}{\beta}\Bigr)^2 dt_b^2 + d\tilde{\rho}^2~,
\ee
after the rescaling $\tilde{\rho} = 2\pi\rho/\beta$ that defines the modular time coordinate $t_b = \beta\tau/(2\pi)$. We work in local null coordinates centered on the Hawking particle at modular time $t_b$. The outgoing null coordinate on the future horizon is given by the Rindler exponential map
\be
v = e^{2\pi t_b/\beta}~.
\ee
In these coordinates the horizon lies at $v\to 0$ as $t_b \to -\infty $, while larger values of $v$ correspond to points farther into the exterior. This coordinate is related to the retarded time, $v_{\rm ret} = t_b - (\beta/2 \pi)\log(2\pi \tilde{\rho}/ \beta)$, by the identity $v=2 \pi \tilde{\rho}/\beta\cdot\exp(\kappa u_{\rm ret})$. We work with the finite the coordinate $u$, which remains regular on the horizon in the limit $\tilde{\rho}\to 0$, while $u_{\rm ret} \to \infty $. The vanishing factor $\tilde{\rho}$ exactly compensates the logarithmic divergence, so that $u$ provides a regular parametrization of points along the future cosmological horizon.

We place the exterior Hawking operator $\phi$ at null position $v'$ in its local frame. Its mirror partner appears at the effective position $\bar{v} = -v_b$, where $v_b = e^{2\pi t_b/\beta}$ labels the creation time of the pair. For a primary operator of conformal dimension $\Delta = m\ell$, the light-cone two-point function on the horizon is universal. Substituting the exponential map into the CFT correlator, yields
\be\label{eq:horizon}
K_0(v'; t_b) = \Bigl( v' + e^{-2\pi t_b / \beta} \Bigr)^{-2\Delta}~.
\ee
Consider the two-point function in the perturbed state
\be
f(z) := \bra\Phi \phi(t_b + z) \, \bar{\phi}(0) \ket \Phi~.
\ee
Analytic continuation properties of the relative modular operator imply that $f(z)$ extends holomorphically into the strip $0 < \Im z < \beta$. Its value on the upper boundary of the strip is given by the same expectation value in which the imaginary-time shift has been implemented by the effective relative modular flow
\be
\label{eq:kms}
f(t+i \beta) = \bra\Phi \bar{\phi}(0) \, \sigma_t^{\Phi|\Psi} \bigl( \phi(t_b + t) \bigr) \ket\Phi~,
\ee
where the flow parameter $t$ is dimensionless. Here the factor $\beta$ appears because the modular Hamiltonian is normalized so that the KMS period equals the inverse background temperature.

The entropy computed earlier in Sec. \ref{sec:araki} is the von Neumann entropy of the perturbed probe state, whose second term is the relative entropy $\Delta S_{\rm rel}(\Phi||\Psi) = -\langle\Phi|\log\Delta_{\Phi|\Psi}|\Phi\rangle$. This entropy measures the distinguishability between the Hawking-pair state and the reference microcanonical TFD. It is therefore essential to use the effective relative flow rather than the ordinary modular flow generated by $\Delta_\Phi$ or $\Delta_\Psi$ individually, which only implement thermal evolution within each separate state. The effective relative flow defined above is the object that correctly compares the two states and extracts the real-time dynamics induced by the Hawking pair.

Because the HKLL modes are free fields, the relative modular flow acts on horizon operators by a null-coordinate shift
\be
\sigma_t^{\Phi|\Psi}(\phi(v')) = \phi\bigl(v' + \delta(t; t_b)\bigr)~.
\ee
Substituting the explicit form of the horizon propagator (\ref{eq:horizon}) into the KMS relation (\ref{eq:kms}) and matching the additive term under the analytic continuation immediately yields
\be
\label{eq:shift}
\delta(t; t_b) = e^{-2\pi t_b / \beta} \bigl(1 - e^{-2\pi t / \beta}\bigr)~.
\ee
Thus the geometric action of the modular flow on the horizon region becomes\footnote{Similarly, the mirror operator $\bar{\phi}$ transforms under the inverse effective flow, so its shift is
\be
\sigma_{-t}^{\Phi|\Psi}(\bar{\phi}(\bar{v})) = \bar{\phi}\Bigl( \bar{v} + e^{-2\pi t_b / \beta} \bigl(e^{2\pi t / \beta}-1\bigr) \Bigr)~.
\ee}
\be
\sigma_t^{\Phi|\Psi}(\phi(v')) = \phi\Bigl( v' + e^{-2\pi t_b / \beta} (1 - e^{-2\pi t / \beta}) \Bigr)~.
\ee

In the near horizon regime ($t_b \to -\infty $), the factor $e^{-2\pi t_b / \beta}$ becomes exponentially large. For a positive flow parameter $t>0$ the term in parenthesis is positive, so the exterior operator experiences a large positive null shift that moves it farther from the horizon, while the mirror partner behind the horizon receives the opposite shift under the inverse flow. The net effect is that the Hawking pair is taken apart across the horizon and their effective null separation increases proportionally to the shift $\delta$. 

This stretching originates from the boost structure of the near-horizon geometry and produces a clear redshift/blueshift asymmetry. The exterior mode still suffers gravitational redshift as it propagates away from the horizon, but the positive shift reduces the net redshift it experiences when observed by a static-patch observer. In contrast, the mirror partner is taken deeper into the interior and experiences a blueshift. In this regime the relation between modular time and the null coordinate is $dt_b \approx \beta dv/2\pi v$, which diverges as $v \to 0$. This implies that the exterior mode remains effectively `frozen' (strongly redshifted) near the horizon until it has propagated a macroscopic affine distance outward, while the interior partner carries high-energy correlations deeper inside. 

As a result, quantum information is not released the instant the pair is created. The exterior quantum mode remains effectively trapped close to the horizon until the accumulated stretch becomes order one. Only then does the redshift/blueshift asymmetry allow the entanglement between the partners to decrease sufficiently for information to become accessible to the static-patch observer. This local backreaction--the effective null shift acting as a small positive-energy shockwave on the horizon--is precisely what enables the static-patch observer to detect the outgoing Hawking particle at infinity.

The exponential growth of the effective null shift 
(\ref{eq:shift}) generated by the relative modular flow is characteristic of chaos. It emerges entirely within the algebraic framework from the properties of the Hawking-pair state and the relative modular operator $\Delta_{\Phi|\Psi}$ acting on the crossed-product Type II$_\infty$ algebra $\tilde{\mathcal{A}}$. Differentiating the dominant term in the shift immediately yields the Lyapunov exponent $\lambda_L = 2\pi/\beta$. The scrambling time is the moment at which this shift becomes non-perturbative
\be
t_{\rm scr} \approx \frac{\beta}{2\pi} \log S~,
\ee
where $S$ is the coarse-grained entropy of the cosmological horizon. Analytic continuation of the horizon correlator into the KMS strip converts its decay into the characteristic early-time exponential growth of the associated commutator squared, which saturates at $t\approx t_{\rm scr}$.

This mechanism directly explains the slow rise of $\Delta S_{\rm bulk}$ near $v=0$. Information carried by the Hawking pair remains ``frozen'' near the horizon until the accumulated stretch generated by the relative modular flow becomes order-one. Only then does the exterior partner escape while the interior mirror is driven deep into the dS region, after which the state again appears thermal to the static-patch observer and the entropy contribution returns to zero at late times. This timescale is precisely the Hayden-Preskill recovery time $t_{\rm scr}$. The pair is created locally near the horizon, but its quantum information only becomes accessible to an exterior observer after a delay of order the scrambling time. In this sense our Hawking pair calculation provides a microscopic, algebraic realization of information transfer for flow geometries—information does not leak out immediately, but is delocalized over the horizon degrees of freedom and released only after maximal scrambling has occurred.

\section{Discussion}  
In this work, we have revisited the question of information transfer across horizons, with a particular focus on when information escapes a cosmological horizon. While powerful algebraic probes have been developed to track the timescale at which information leaves an evaporating black hole, comparatively little is known about the analogous process for de Sitter horizons. Cosmological horizons are observer-dependent and exhibit distinctive thermodynamics, making them a sharper theoretical challenge. To address this, we have studied two-dimensional flow geometries that interpolate smoothly between an asymptotic AdS$_2$ boundary and a dS$_2$ static patch. These geometries provide a controlled holographic arena in which the physics of the cosmological horizon can be probed using the standard tools of AdS/CFT while retaining the essential features of de Sitter space.

Using the algebraic formalism of von Neumann algebras, we have constructed the centaur-algebra of observables and modeled a Hawking pair as a probe state. By acting with local operators in the Euclidean theory and performing the appropriate Lorentzian continuation, we prepared states that capture both the near-horizon and far-away regimes of the pair. Promoting the algebra to a crossed-product Type II$_\infty$ factor allowed us to define a rigorous notion of entropy difference between the reference thermofield-double state and the state containing the Hawking pair. The resulting entropy curve is an inverse mini-Page curve: it decreases from zero, reaches a minimum at Euclidean time separation $\tau \approx \beta/8$, and then increases again. This algebraic mini-Page curve supplies a sharp, quantitative answer to the question of when information escapes a de Sitter horizon and opens a new window onto the quantum features of cosmological spacetimes.

The Hawking-pair construction is intimately connected to our earlier work on flow geometries and partially entangled thermal states (PETS) in the microcanonical ensemble \cite{Espindola:2025wjf}. The Hawking-pair state studied here is the simplest probe that already encodes the essential information-transfer mechanism; more general PETS states \cite{Goel:2018ubv,Gesteau:2025obm,Kudler-Flam:2025cki,Liu:2025ikq} can be viewed as coherent superpositions of such pairs. The techniques developed in this paper therefore provide a microscopic foundation for a broader algebraic treatment of quantum information in de Sitter and near-de Sitter spacetimes.

Finally, our results on the relative modular operator and the associated effective modular flow already exhibit clear signatures of chaos. In Sec. \ref{sec:Sgen}, we derived an effective null-coordinate shift on horizon operators whose exponential growth yields a Lyapunov exponent $  \lambda_L = 2\pi/\beta  $ and identifies the scrambling time as the moment when information carried by the pair becomes accessible to a static-patch observer. It would be very interesting to study the chaos properties of flow-geometry horizons in greater depth using the recently developed algebraic framework for gravitational scrambling introduced by \cite{Penington:2025hrc}. The centaur-algebra constructed in this work provides a natural and concrete setting in which to apply that framework, potentially allowing a precise characterization of out-of-time-order correlators, modular scrambling \cite{DeBoer:2019kdj}, and operator growth near cosmological horizons. We leave this direction for future investigation.

\section*{Acknowledgments}
It is a pleasure to thank Bart{\l}omiej Czech, Bik Soon Sia, Roberto Emparan, Viktor Jahnke, Herman Verlinde and Zhenbin Yang for illuminating discussions. RE thanks the Institute of Cosmos Sciences of the University of Barcelona (ICCUB) where part of this work was carried out. RE thanks the organizers of the workshops ``Holo-Asia 2026'' and ``Quantum Information in Quantum Gravity 2026'' (QIQG 2026). RE is supported by the Dushi Zhuanxiang Fellowship and acknowledges a Shuimu Scholarship as part of the Shuimu Tsinghua Scholar Program. SM is supported in part by MEXT KAKENHI Grant Number 21H05187.

\appendix

\section{Euclidean two-point function in JT gravity} \label{App:JTtwopt}

In this appendix, we briefly explain the derivation of the Euclidean two-point function in JT gravity.
The metric of the finite-temperature black hole solution is
\begin{align}
ds^2 = f_{BH}(r) \, d\tau^2 + \frac{1}{f_{BH}(r)} \, dr^2 , \qquad 
f_{BH}(r) = - \frac{4\pi^2 \ell^2}{\beta^2} + \frac{r^2}{\ell^2}.
\end{align}
The energy of the particle is similarly defined as \eqref{eq:geocon}
\be
\mathcal{E} = f_{BH}(r(s)) \dot{\tau}(s) ~ ,
\ee
and the constraint equation for the geodesic \eqref{eq:geofix} now becomes
\begin{align}
\frac{\mathcal{E}^2}{f_{BH}(r(s))} + \frac{\dot{r}(s)^2}{f_{BH}(r(s))} = 1
\quad \Longleftrightarrow \quad
\dot{r}(s)^2 = -\mathcal{E}^2 - \frac{4\pi^2 \ell^2}{\beta^2} + \frac{r^2}{\ell^2}.
\end{align}
The solution is
\begin{align}
r(s) = \ell \sqrt{\mathcal{E}^2 + \frac{4\pi^2 \ell^2}{\beta^2}} \cosh \left( \frac{s}{\ell} \right), \hspace{0.4cm}\\
\tau(s) =  \frac{ \beta }{2\pi} \arctan \left[ \frac{2\pi \ell }{\mathcal{E} \beta } \tanh\left( \frac{s}{\ell} \right) \right].
\end{align}

The important difference from the centaur gravity case is that the functional form of $\tau(s)$ now involves an $\arctan$ instead of an $\operatorname{arctanh}$.
Due to this difference, the relation between $\mathcal{E}$ and $\Delta \tau$ becomes
\begin{align}
\Delta \tau 
= \frac{\beta}{\pi} \arctan \left[ \frac{2\pi \ell}{\mathcal{E} \beta} \right]
\qquad \Longrightarrow \qquad \mathcal{E} = \frac{2\pi \ell}{\beta} \frac{1}{\tan( \frac{\pi \Delta \tau}{\beta} )}.
\end{align}
Thus,
\begin{align}
r(s)= \frac{2\pi \ell^2}{\beta}  \frac{1}{\sin ( \frac{\pi \Delta \tau}{\beta} )}  \cosh (\frac{s}{\ell}) ~ , \hspace{1.9cm} \\
\tau(s) =  \frac{ \beta }{2\pi} \arctan \left[ \tan( \frac{\pi \Delta \tau}{\beta} ) \tanh\left( \frac{s}{\ell} \right) \right] ~ .
\end{align}
A sine function appears instead of the hyperbolic sine in the centaur case.

\noindent
Next, let us compute the on-shell action with a little more detail than the main text. As noted in the main text, the action we use is \eqref{eq:totalaction} - \eqref{eq:counteraction};
\begin{align}
I^{\rm p}[\gamma] = I_{\rm bare}^{\rm p}[\gamma] + I_{\rm ct}^{\rm p}(\partial \gamma) ~ ,   \hspace{6cm} \notag \\
I_{\rm bare}^{\rm p}[\gamma] = m \int_{\gamma} \sqrt{h} ds ~~~~~ , 
I_{\rm ct}^{\rm p}(\partial \gamma) = \left. - m \ell \log \left( \frac{2 \ell \Phi}{\mathcal{C}} \right) \right|_{\partial \gamma} ~ . \notag 
\end{align}
Here, the dilaton configuration is given by
\be
\Phi(r) = \widetilde{\Phi}_{b} \frac{r}{\ell} ~  .
\ee
Let us introduce the large cutoff $r=r_{\infty}$ and corresponding proper time cutoff $s_{\infty}$ by
\be
r_{\infty} = r(s_{\infty}) ~ .
\ee
For large $r_{\infty}$ (thus large $s_{\infty}$), it reduces to the following relation;
\be
r_{\infty} \simeq \frac{\pi \ell^2}{\beta}  \frac{1}{\sin ( \frac{\pi \Delta \tau}{\beta} )}  \exp(\frac{s_\infty}{\ell}) ~ .
\ee
Using $s_{\infty}$ and the above relation, the on-shell value of the bare action $I^{\rm p}_{\rm bare}|_{\rm on-shell}$ and the counterterm $I^{\rm p}_{ct}(\partial \gamma)$ are given by
\begin{align}
I^{\rm p}_{\rm bare}|_{\rm on-shell} = 2 m s_{\infty} ~ , \hspace{8cm} \\
I^{\rm p}_{ct}(\partial \gamma) \simeq -2m \ell \log\left( \frac{2 \pi \ell^2 \widetilde{\Phi}_{b} }{\mathcal{C} \beta} \frac{1}{\sin( \frac{\pi \Delta \tau}{\beta} )} e^{\frac{s_{\infty}}{\ell}} \right) \hspace{3.9cm} \notag \\
= -2 m s_{\infty}  - 2 m \ell \log\left( \frac{2 \pi \ell^2 \widetilde{\Phi}_{b} }{\mathcal{C} \beta} \frac{1}{\sin( \frac{\pi \Delta \tau}{\beta} )}  \right) ~~~ ({\rm for ~ } s_{\infty} \gg \ell) ~ .
\end{align}
Therefore, when they are summed, $I^{\rm p}_{\rm bare}|_{\rm on-shell}$ and the first term of $I^{\rm p}_{ct}(\partial \gamma)$ are canceled, and the total on-shell action becomes
\be
I^{\rm p}|_{\rm on-shell} =  - 2 m \ell \log\left( \frac{2 \pi \ell^2 \widetilde{\Phi}_{b} }{\mathcal{C} \beta} \frac{1}{\sin( \frac{\pi \Delta \tau}{\beta} )}  \right) ~ .
\ee
Thus, the correlation function is
\begin{align}
G_{\beta}^{\rm (JT)}(\Delta \tau) \simeq e^{-I^{\rm p}|_{\rm on-shell}}
= \left( \frac{2 \pi \ell^2 \widetilde{\Phi}_{b} }{\mathcal{C} \beta} \frac{1}{\sin( \frac{\pi \Delta \tau}{\beta} )}  \right)^{2m\ell}.
\end{align}
Using this two-point function, $ \Delta S_{\rm grav}$ in JT gravity is computed as
\begin{align}
\Delta S_{\rm grav} & = -n \partial_n \log \left( G_{n \beta}^{\rm(JT)} (y)^2\right) \lvert_{n=1} ~ , \\
& = 4 m \ell \left( 1 - \frac{\pi y }{\beta} \cot(\frac{\pi y}{\beta} ) \right) ~ .
\end{align}

\section{Crossed product}
\label{App:crossed}

In this appendix we review the crossed product construction. This construction is central to the algebraic framework of the paper. It supplies a well-defined trace on the enlarged algebra and thereby allows a rigorous definition of gravitational and relative entropies. To make the construction more intuitive, we first recall the analogous operation in group theory--the semidirect product--and then illustrate the procedure in detail in the familiar setting of an AdS black hole, following the approach of \cite{Witten:2021unn}.

\subsection{Group theory and algebra}
Let $H$ and $K$ be two groups, and let $\phi: K \rightarrow Aut(H)$ be an action of $K$ on $H$ by group automorphisms. The semidirect product $H \rtimes_{\varphi} K$ is defined on the set $H \times K$ with the group multiplication
\be
(h,k)(h',k') = \bigl(h \cdot \phi_k(h'),~kk' \bigr)~.
\ee
In this larger group, the subgroup  $H \times \{e\}$ is always normal. By contrast, the subgroup $\{e\} \times K$ is normal if and only if the action $\phi$ is trivial. When the action is trivial, the semidirect product reduces to the ordinary direct product $H \times K$.

This construction provides a useful analogy for what follows. One starts with a group $H$ and consistently adjoins an action of another group $K$ by automorphisms. The resulting object is larger and genuinely new whenever the action is non-trivial.

The crossed product of a von Neumann algebra by a group action is the operator-algebraic anlogue of the semidirect product. Let $\cA$  be a von Neumann algebra acting on a Hilbert space $\cH$ arising from the GNS construction associated with a faithful normal state. Suppose a group $G$ acts on $\cA$ by automorphisms $  \alpha_g  $. In order to implement this action, we enlarge the Hilbert space to 
\be
\hat{\mathcal{H}} = \mathcal{H} \otimes L^2(G)~,
\ee
where $L^2(G)$ is the space of square-integrable functions on $G$ with respect to the normalized Haar measure. Vectors in $\hat{\mathcal{H}}$ can be viewed as functions $g \mapsto \ket{\xi(g)} \in \mathcal{H}$.

The crossed-product algebra $\mathcal{A} \rtimes_\alpha G$ is the von Neumann algebra generated by two sets of operators acting on $\hat{\mathcal{H}}$. The original algebra $\cA$ acts by 
\be
(a \cdot \xi)(g) = \alpha_{g^{-1}}(a) \xi(g)~,
\ee
while the group $G$ acts by left translation
\be
(U_h \xi)(g) = \xi(h^{-1} g)~.
\ee
The resulting algebra $\cA \rtimes_\alpha G  $ acts on the enlarged space $\hat{\mathcal{H}}$ and does not depend on the particular choice of the original GNS Hilbert space $\mathcal{H}$.

A equivalent and often helpful way to characterize the crossed product is through an invariance condition. One can start with the larger algebra $\mathcal{A} \overline{\otimes} B(L^2(G))$ and then select the operators that are invariant under a natural action of $G$. Specifically, the crossed product consists of those elements $x$ satisfying 
\be
U_h x U_h^{-1} = x~, \quad \text{for all } h \in G~,
\ee
where $U_h$ implements the automorphism  $\alpha_h$ on the first factor and acts by translation on the second factor. In this sense, the crossed product is obtained by ``adjoining the group action and then imposing invariance,'' in close analogy with how the semidirect product incorporates an action by automorphisms. When the action of $G$ on $\cA$ is outer, the crossed product algebra $\cA \rtimes G  $ is a genuinely larger algebra. In the case relevant to this paper, this procedure converts a Type III$_1  $ algebra into a Type II$_\infty$ factor that admits a faithful normal semifinite trace.

\subsection{AdS black holes}
Consider an AdS black hole in thermal equilibrium at inverse temperature $\beta$, whose holographic dual is the thermofield double state $\ket{\Psi_{\TFD}}$. Let $\cA_{r,0}$ denote the von Neumann algebra of observables the right exterior region; this is a Type III$_1$ factor. We enlarge this algebra by means of the crossed product

Let $T$ a self-adjoint operator on the Hilbert space that generates a one-parameter group of automorphisms of $\cA_{r,0}$
\be
\label{eq:automorphism}
e^{i T s} a e^{-iTs} \in \cA_{r,0}~, ~~~~ \forall a \in \cA_{r,0} \quad {\rm and} \quad s\in \mathbb{R}~.
\ee
The crossed product algebra $\cA_{r,0} \rtimes \mathbb{R}$ acts on the extended Hilbert space 
\be
\hat{\cH}=\cH_{\TFD} \otimes L^2(\mathbb{R})~,
\ee
where $L^2(\mathbb{R})$ is the space of square-integrable functions of a real variable $X$. Concretely, the crossed product algebra is generated by operators of the form 
\be
a e^{isT} \otimes e^{isX}~, ~ {\rm with}~ a \in \cA_{r,0} \quad {\rm and} \quad s \in \mathbb{R}~.
\ee

An automorphism is said to be inner if the implementing operator $e^{isT}$ already belongs to $\cA_{r,0}$; otherwise it is outer. If the automorphism is inner, adjoining $T+X$ simply produces a direct product and yields nothing new. When the automorphism is outer, however, the crossed product $\cA_{r,0} \rtimes \mathbb{R}$ becomes a factor. In particular, when $T$ is chosen to be the modular Hamiltonian $\hat{h}$ of the thermofield-double state (which generates the modular automorphism group), the resulting algebra
\be
\cA_R = \cA_{r,0} \rtimes \mathbb{R}~.
\ee
is a Type II$_\infty$ factor. This algebra admits a faithful normal semifinite trace, which in turn permits a well-defined notion of entropy.

It is convenient to work with classical-quantum states of the form 
\be 
\h{\Psi} = \Psi_{\TFD} \otimes g^{1/2}(X)~, \quad {\rm with} \quad g^{1/2}(X) \in L^2(\mathbb{R})~.
\ee 
Our goal is to construct the modular operator $\h{\Delta}_{\h{\Psi}}$ associated with this state on the crossed-product algebra. Consider two generic operators 
\be
\h\a=a e^{is(\h h + X)}~, \quad ~ \h\b=b e^{it(\h h + X)}~,
\ee
with $a, b \in \cA_{r,0}$. Their expectation value in the state $\ket{\h\Psi}$ reads
\be 
\bra{\h{\Psi}}\h\a \h\b \ket{{\h{\Psi}}} =\int dX ~ g(X) e^{i(t+s) X} \bra{\Psi} \a \b_s \ket{\Psi}~,
\ee
where $\b_s=e^{is \h h} \b e^{-is \h h}$. Introducing the Fourier transform $\tilde{g}(w)$ of $g(X)$ and using the KMS condition
\be
\bra{\Psi} \a_u \b \ket{\Psi} = \bra{\Psi} \b \a_{u+i} \ket{\Psi}~,
\ee
we obtain
\be
\bra{\h\Psi} \h \a \h \b \ket{\h \Psi} = \int dX g(X) \int dw \bra{\Psi} b \tilde{g}(w) \Delta_\Psi e^{-i w (\h h + X)} a \ket{\Psi}~.
\ee
Defining the operators 
\be
K = e^{-(h + X)} g( h + X) ~~~ {\rm and} ~~~ \t K = \frac{e^X}{g(X)}~.
\ee
we arrive at the desired modular operator on the crossed-product algebra
\be
\bra{\h{\Psi}}\h\a\h\b \ket{{\h{\Psi}}} = \bra{\h{\Psi}} \h\b \h\Delta_{\h\Psi} \h\a \ket{{\h{\Psi}}}~, \h \Delta_{\h \Psi} = \tilde{K}K ~.
\ee
Because the operator $\h\Delta_{\h \Psi}$ factorizes, the modular automorphism it generates is inner on $\cA_R$. This completes the construction of the Type II$_\infty$ algebra and its associated modular operator for the AdS black hole case.

\bibliographystyle{jhep}
\bibliography{ref.bib}

@article{Aguilar-Gutierrez:2023odp,
    author = "Aguilar-Gutierrez, Sergio E. and Bahiru, Eyoab and Esp{\'\i}ndola, Ricardo",
    title = "{The centaur-algebra of observables}",
    eprint = "2307.04233",
    archivePrefix = "arXiv",
    primaryClass = "hep-th",
    doi = "10.1007/JHEP03(2024)008",
    journal = "JHEP",
    volume = "03",
    pages = "008",
    year = "2024"
}

@article{Almheiri:2019psf,
    author = "Almheiri, Ahmed and Engelhardt, Netta and Marolf, Donald and Maxfield, Henry",
    title = "{The entropy of bulk quantum fields and the entanglement wedge of an evaporating black hole}",
    eprint = "1905.08762",
    archivePrefix = "arXiv",
    primaryClass = "hep-th",
    doi = "10.1007/JHEP12(2019)063",
    journal = "JHEP",
    volume = "12",
    pages = "063",
    year = "2019"
}

@article{Almheiri:2019hni,
    author = "Almheiri, Ahmed and Mahajan, Raghu and Maldacena, Juan and Zhao, Ying",
    title = "{The Page curve of Hawking radiation from semiclassical geometry}",
    eprint = "1908.10996",
    archivePrefix = "arXiv",
    primaryClass = "hep-th",
    doi = "10.1007/JHEP03(2020)149",
    journal = "JHEP",
    volume = "03",
    pages = "149",
    year = "2020"
}

@article{Almheiri:2019qdq,
    author = "Almheiri, Ahmed and Hartman, Thomas and Maldacena, Juan and Shaghoulian, Edgar and Tajdini, Amirhossein",
    title = "{Replica Wormholes and the Entropy of Hawking Radiation}",
    eprint = "1911.12333",
    archivePrefix = "arXiv",
    primaryClass = "hep-th",
    doi = "10.1007/JHEP05(2020)013",
    journal = "JHEP",
    volume = "05",
    pages = "013",
    year = "2020"
}

@article{Anninos:2017hhn,
    author = "Anninos, Dionysios and Hofman, Diego M.",
    title = "{Infrared Realization of dS$_2$ in AdS$_2$}",
    eprint = "1703.04622",
    archivePrefix = "arXiv",
    primaryClass = "hep-th",
    doi = "10.1088/1361-6382/aab143",
    journal = "Class. Quant. Grav.",
    volume = "35",
    number = "8",
    pages = "085003",
    year = "2018"
}

@article{Anninos:2018svg,
    author = "Anninos, Dionysios and Galante, Dami{\'a}n A. and Hofman, Diego M.",
    title = "{De Sitter horizons {\&} holographic liquids}",
    eprint = "1811.08153",
    archivePrefix = "arXiv",
    primaryClass = "hep-th",
    doi = "10.1007/JHEP07(2019)038",
    journal = "JHEP",
    volume = "07",
    pages = "038",
    year = "2019"
}

@article{Almheiri:2019yqk,
    author = "Almheiri, Ahmed and Mahajan, Raghu and Maldacena, Juan",
    title = "{Islands outside the horizon}",
    eprint = "1910.11077",
    archivePrefix = "arXiv",
    primaryClass = "hep-th",
    month = "10",
    year = "2019"
}

@article{Antonini:2025sur,
    author = "Antonini, Stefano and Chen, Chang-Han and Maxfield, Henry and Penington, Geoff",
    title = "{An apologia for islands}",
    eprint = "2506.04311",
    archivePrefix = "arXiv",
    primaryClass = "hep-th",
    doi = "10.1007/JHEP10(2025)034",
    journal = "JHEP",
    volume = "10",
    pages = "034",
    year = "2025"
}

@article{Brown:2019rox,
    author = "Brown, Adam R. and Gharibyan, Hrant and Penington, Geoff and Susskind, Leonard",
    title = "{The Python{\textquoteright}s Lunch: geometric obstructions to decoding Hawking radiation}",
    eprint = "1912.00228",
    archivePrefix = "arXiv",
    primaryClass = "hep-th",
    doi = "10.1007/JHEP08(2020)121",
    journal = "JHEP",
    volume = "08",
    pages = "121",
    year = "2020"
}

@article{Bekenstein:1972tm,
    author = "Bekenstein, J. D.",
    title = "{Black holes and the second law}",
    doi = "10.1007/BF02757029",
    journal = "Lett. Nuovo Cim.",
    volume = "4",
    pages = "737--740",
    year = "1972"
}

@article{Bekenstein:1973ur,
    author = "Bekenstein, Jacob D.",
    title = "{Black holes and entropy}",
    doi = "10.1103/PhysRevD.7.2333",
    journal = "Phys. Rev. D",
    volume = "7",
    pages = "2333--2346",
    year = "1973"
}

@article{Bousso:2023kdj,
    author = "Bousso, Raphael and Penington, Geoff",
    title = "{Islands far outside the horizon}",
    eprint = "2312.03078",
    archivePrefix = "arXiv",
    primaryClass = "hep-th",
    doi = "10.1007/JHEP11(2024)164",
    journal = "JHEP",
    volume = "11",
    pages = "164",
    year = "2024"
}

@article{Bousso:2022gth,
    author = "Bousso, Raphael and Wildenhain, Elizabeth",
    title = "{Islands in closed and open universes}",
    eprint = "2202.05278",
    archivePrefix = "arXiv",
    primaryClass = "hep-th",
    doi = "10.1103/PhysRevD.105.086012",
    journal = "Phys. Rev. D",
    volume = "105",
    number = "8",
    pages = "086012",
    year = "2022"
}

@article{Chandrasekaran:2022eqq,
    author = "Chandrasekaran, Venkatesa and Penington, Geoff and Witten, Edward",
    title = "{Large N algebras and generalized entropy}",
    eprint = "2209.10454",
    archivePrefix = "arXiv",
    primaryClass = "hep-th",
    doi = "10.1007/JHEP04(2023)009",
    journal = "JHEP",
    volume = "04",
    pages = "009",
    year = "2023"
}

@article{Chandrasekaran:2022cip,
    author = "Chandrasekaran, Venkatesa and Longo, Roberto and Penington, Geoff and Witten, Edward",
    title = "{An algebra of observables for de Sitter space}",
    eprint = "2206.10780",
    archivePrefix = "arXiv",
    primaryClass = "hep-th",
    doi = "10.1007/JHEP02(2023)082",
    journal = "JHEP",
    volume = "02",
    pages = "082",
    year = "2023"
}

@article{Connes1973,
author = {Connes, Alain},
journal = {Annales scientifiques de l'École Normale Supérieure},
language = {fre},
number = {2},
pages = {133-252},
publisher = {Elsevier},
title = {Une classification des facteurs de type $\{\rm III\}$},
url = {http://eudml.org/doc/81916},
volume = {6},
year = {1973},
}

@article{Chen:2019iro,
    author = "Chen, Yiming",
    title = "{Pulling Out the Island with Modular Flow}",
    eprint = "1912.02210",
    archivePrefix = "arXiv",
    primaryClass = "hep-th",
    doi = "10.1007/JHEP03(2020)033",
    journal = "JHEP",
    volume = "03",
    pages = "033",
    year = "2020"
}

@article{Chen:2025tbh,
    author = "Chen, Bin and Xu, Jie",
    title = "{An algebra for covariant observers in de Sitter space}",
    eprint = "2511.00622",
    archivePrefix = "arXiv",
    primaryClass = "hep-th",
    month = "11",
    year = "2025"
}

@article{DeBoer:2019kdj,
    author = "De Boer, Jan and Lamprou, Lampros",
    title = "{Holographic Order from Modular Chaos}",
    eprint = "1912.02810",
    archivePrefix = "arXiv",
    primaryClass = "hep-th",
    doi = "10.1007/JHEP06(2020)024",
    journal = "JHEP",
    volume = "06",
    pages = "024",
    year = "2020"
}

@article{Espindola:2025wjf,
    author = "Esp{\'\i}ndola, Ricardo and Miyashita, Shoichiro",
    title = "{Flow-geometry microstates}",
    eprint = "2510.18901",
    archivePrefix = "arXiv",
    primaryClass = "hep-th",
    doi = "10.1007/JHEP06(2026)021",
    journal = "JHEP",
    volume = "06",
    pages = "021",
    year = "2026"
}

@article{Espindola:2026ekv,
    author = "Esp{\'\i}ndola, Ricardo and Ali, Ahmed Farag",
    title = "{Spectral Admissibility of Real Observers in Euclidean de Sitter Gravity}",
    eprint = "2605.30423",
    archivePrefix = "arXiv",
    primaryClass = "hep-th",
    month = "5",
    year = "2026"
}

@article{Espindola:2022fqb,
    author = "Esp{\'\i}ndola, Ricardo and Najian, Bahman and Nikolakopoulou, Dora",
    title = "{Islands in FRW Cosmologies}",
    eprint = "2203.04433",
    archivePrefix = "arXiv",
    primaryClass = "hep-th",
    month = "3",
    year = "2022"
}

@article{Espindola:2025ons,
    author = "Esp{\'\i}ndola, Ricardo and Jahnke, Viktor and Kim, Keun-Young",
    title = "{Islands and traversable wormholes}",
    eprint = "2510.21985",
    archivePrefix = "arXiv",
    primaryClass = "hep-th",
    month = "10",
    year = "2025"
}

@article{Faulkner:2024gst,
    author = "Faulkner, Thomas and Speranza, Antony J.",
    title = "{Gravitational algebras and the generalized second law}",
    eprint = "2405.00847",
    archivePrefix = "arXiv",
    primaryClass = "hep-th",
    doi = "10.1007/JHEP11(2024)099",
    journal = "JHEP",
    volume = "11",
    pages = "099",
    year = "2024"
}

@article{Gesteau:2025obm,
    author = "Gesteau, Elliott",
    title = "{A no-go theorem for large $N$ closed universes}",
    eprint = "2509.14338",
    archivePrefix = "arXiv",
    primaryClass = "hep-th",
    reportNumber = "MIT-CTP/5927",
    month = "9",
    year = "2025"
}

@article{Gibbons:1977mu,
    author = "Gibbons, G. W. and Hawking, S. W.",
    title = "{Cosmological Event Horizons, Thermodynamics, and Particle Creation}",
    doi = "10.1103/PhysRevD.15.2738",
    journal = "Phys. Rev. D",
    volume = "15",
    pages = "2738--2751",
    year = "1977"
}

@article{Goel:2018ubv,
    author = "Goel, Akash and Lam, Ho Tat and Turiaci, Gustavo J. and Verlinde, Herman",
    title = "{Expanding the Black Hole Interior: Partially Entangled Thermal States in SYK}",
    eprint = "1807.03916",
    archivePrefix = "arXiv",
    primaryClass = "hep-th",
    doi = "10.1007/JHEP02(2019)156",
    journal = "JHEP",
    volume = "02",
    pages = "156",
    year = "2019"
}

@article{Gubser:1998bc,
    author = "Gubser, S. S. and Klebanov, Igor R. and Polyakov, Alexander M.",
    title = "{Gauge theory correlators from noncritical string theory}",
    eprint = "hep-th/9802109",
    archivePrefix = "arXiv",
    reportNumber = "PUPT-1767",
    doi = "10.1016/S0370-2693(98)00377-3",
    journal = "Phys. Lett. B",
    volume = "428",
    pages = "105--114",
    year = "1998"
}

@article{Gesteau:2024dhj,
    author = "Gesteau, Elliott and Santilli, Leonardo",
    title = "{Explicit large $N$ von Neumann algebras from matrix models}",
    eprint = "2402.10262",
    archivePrefix = "arXiv",
    primaryClass = "hep-th",
    doi = "10.4310/atmp.241031230051",
    journal = "Adv. Theor. Math. Phys.",
    volume = "28",
    number = "7",
    pages = "2245--2429",
    year = "2024"
}

@article{Geng:2021wcq,
    author = "Geng, Hao and Nomura, Yasunori and Sun, Hao-Yu",
    title = "{Information paradox and its resolution in de Sitter holography}",
    eprint = "2103.07477",
    archivePrefix = "arXiv",
    primaryClass = "hep-th",
    doi = "10.1103/PhysRevD.103.126004",
    journal = "Phys. Rev. D",
    volume = "103",
    number = "12",
    pages = "126004",
    year = "2021"
}

@article{Hawking:1975vcx,
    author = "Hawking, S. W.",
    editor = "Gibbons, G. W. and Hawking, S. W.",
    title = "{Particle Creation by Black Holes}",
    doi = "10.1007/BF02345020",
    journal = "Commun. Math. Phys.",
    volume = "43",
    pages = "199--220",
    year = "1975",
    note = "[Erratum: Commun.Math.Phys. 46, 206 (1976)]"
}

@article{Hawking:1976ra,
    author = "Hawking, S. W.",
    title = "{Breakdown of Predictability in Gravitational Collapse}",
    doi = "10.1103/PhysRevD.14.2460",
    journal = "Phys. Rev. D",
    volume = "14",
    pages = "2460--2473",
    year = "1976"
}

@article{Hartman:2020khs,
    author = "Hartman, Thomas and Jiang, Yikun and Shaghoulian, Edgar",
    title = "{Islands in cosmology}",
    eprint = "2008.01022",
    archivePrefix = "arXiv",
    primaryClass = "hep-th",
    doi = "10.1007/JHEP11(2020)111",
    journal = "JHEP",
    volume = "11",
    pages = "111",
    year = "2020"
}

@article{Jensen:2023yxy,
    author = "Jensen, Kristan and Sorce, Jonathan and Speranza, Antony J.",
    title = "{Generalized entropy for general subregions in quantum gravity}",
    eprint = "2306.01837",
    archivePrefix = "arXiv",
    primaryClass = "hep-th",
    doi = "10.1007/JHEP12(2023)020",
    journal = "JHEP",
    volume = "12",
    pages = "020",
    year = "2023"
}

@article{Liu:2025ikq,
    author = "Liu, Hong",
    title = "{''Filtering'' CFTs at large N: Euclidean Wormholes, Closed Universes, and Black Hole Interiors}",
    eprint = "2512.13807",
    archivePrefix = "arXiv",
    primaryClass = "hep-th",
    reportNumber = "MIT-CTP/5970",
    month = "12",
    year = "2025"
}

@article{Leutheusser:2021frk,
    author = "Leutheusser, Samuel Aaron Wehlau and Liu, Hong",
    title = "{Emergent Times in Holographic Duality}",
    eprint = "2112.12156",
    archivePrefix = "arXiv",
    primaryClass = "hep-th",
    reportNumber = "MIT-CTP/5382",
    doi = "10.1103/PhysRevD.108.086020",
    journal = "Phys. Rev. D",
    volume = "108",
    number = "8",
    pages = "086020",
    year = "2023"
}

@inproceedings{Liu:2025krl,
    author = "Liu, Hong",
    title = "{Lectures on entanglement, von Neumann algebras, and emergence of spacetime}",
    booktitle = "{Theoretical Advanced Study Institute in Elementary Particle Physics 2023}: {Aspects of Symmetry}",
    eprint = "2510.07017",
    archivePrefix = "arXiv",
    primaryClass = "hep-th",
    reportNumber = "MIT-CTP/5938",
    month = "10",
    year = "2025"
}

@article{Maldacena:1997re,
    author = "Maldacena, Juan Martin",
    title = "{The Large $N$ limit of superconformal field theories and supergravity}",
    eprint = "hep-th/9711200",
    archivePrefix = "arXiv",
    reportNumber = "HUTP-97-A097, HUTP-98-A097",
    doi = "10.4310/ATMP.1998.v2.n2.a1",
    journal = "Adv. Theor. Math. Phys.",
    volume = "2",
    pages = "231--252",
    year = "1998"
}

@article{Page:1993df,
    author = "Page, Don N.",
    title = "{Average entropy of a subsystem}",
    eprint = "gr-qc/9305007",
    archivePrefix = "arXiv",
    reportNumber = "ALBERTA-THY-22-93",
    doi = "10.1103/PhysRevLett.71.1291",
    journal = "Phys. Rev. Lett.",
    volume = "71",
    pages = "1291--1294",
    year = "1993"
}

@inproceedings{Page:1993up,
    author = "Page, Don N.",
    title = "{Black hole information}",
    booktitle = "{5th Canadian Conference on General Relativity and Relativistic Astrophysics (5CCGRRA)}",
    eprint = "hep-th/9305040",
    archivePrefix = "arXiv",
    reportNumber = "ALBERTA-THY-23-93",
    month = "5",
    year = "1993"
}

@article{Page:1993wv,
    author = "Page, Don N.",
    title = "{Information in black hole radiation}",
    eprint = "hep-th/9306083",
    archivePrefix = "arXiv",
    reportNumber = "ALBERTA-THY-24-93",
    doi = "10.1103/PhysRevLett.71.3743",
    journal = "Phys. Rev. Lett.",
    volume = "71",
    pages = "3743--3746",
    year = "1993"
}

@article{Penington:2019npb,
    author = "Penington, Geoffrey",
    title = "{Entanglement Wedge Reconstruction and the Information Paradox}",
    eprint = "1905.08255",
    archivePrefix = "arXiv",
    primaryClass = "hep-th",
    doi = "10.1007/JHEP09(2020)002",
    journal = "JHEP",
    volume = "09",
    pages = "002",
    year = "2020"
}

@article{Penington:2019kki,
    author = "Penington, Geoff and Shenker, Stephen H. and Stanford, Douglas and Yang, Zhenbin",
    title = "{Replica wormholes and the black hole interior}",
    eprint = "1911.11977",
    archivePrefix = "arXiv",
    primaryClass = "hep-th",
    doi = "10.1007/JHEP03(2022)205",
    journal = "JHEP",
    volume = "03",
    pages = "205",
    year = "2022"
}

@article{Penington:2025hrc,
    author = "Penington, Geoff and Tabor, Elisa",
    title = "{The algebraic structure of gravitational scrambling}",
    eprint = "2508.21062",
    archivePrefix = "arXiv",
    primaryClass = "hep-th",
    month = "8",
    year = "2025"
}

@article{Chen:2024rpx,
    author = "Chen, Chang-Han and Penington, Geoff",
    title = "{A clock is just a way to tell the time: gravitational algebras in cosmological spacetimes}",
    eprint = "2406.02116",
    archivePrefix = "arXiv",
    primaryClass = "hep-th",
    month = "6",
    year = "2024"
}

@article{Penington:2023dql,
    author = "Penington, Geoff and Witten, Edward",
    title = "{Algebras and States in JT Gravity}",
    eprint = "2301.07257",
    archivePrefix = "arXiv",
    primaryClass = "hep-th",
    month = "1",
    year = "2023"
}

@article{Susskind:1994vu,
    author = "Susskind, Leonard",
    title = "{The World as a hologram}",
    eprint = "hep-th/9409089",
    archivePrefix = "arXiv",
    reportNumber = "SU-ITP-94-33",
    doi = "10.1063/1.531249",
    journal = "J. Math. Phys.",
    volume = "36",
    pages = "6377--6396",
    year = "1995"
}

@article{Speranza:2025joj,
    author = "Speranza, Antony J.",
    title = "{An intrinsic cosmological observer}",
    eprint = "2504.07630",
    archivePrefix = "arXiv",
    primaryClass = "hep-th",
    doi = "10.1088/1361-6382/ae134c",
    journal = "Class. Quant. Grav.",
    volume = "42",
    number = "21",
    pages = "215023",
    year = "2025"
}

@book{Takesaki:1970aki,
    author = "Takesaki, M.",
    title = "{Tomita's Theory of Modular Hilbert Algebras and its Applications}",
    doi = "10.1007/bfb0065832",
    publisher = "Springer-Verlag",
    series = "Lecture Notes in Mathematics",
    year = "1970"
}

@article{tHooft:1993dmi,
    author = "'t Hooft, Gerard",
    title = "{Dimensional reduction in quantum gravity}",
    eprint = "gr-qc/9310026",
    archivePrefix = "arXiv",
    reportNumber = "THU-93-26",
    journal = "Conf. Proc. C",
    volume = "930308",
    pages = "284--296",
    year = "1993"
}

@article{Verlinde:2022xkw,
    author = "Verlinde, Herman",
    title = "{On the Quantum Information Content of a Hawking Pair}",
    eprint = "2210.08306",
    archivePrefix = "arXiv",
    primaryClass = "hep-th",
    month = "10",
    year = "2022"
}

@article{Kudler-Flam:2025cki,
    author = "Kudler-Flam, Jonah and Witten, Edward",
    title = "{Emergent Mixed States for Baby Universes and Black Holes}",
    eprint = "2510.06376",
    archivePrefix = "arXiv",
    primaryClass = "hep-th",
    month = "10",
    year = "2025"
}

@article{Witten:1998qj,
    author = "Witten, Edward",
    title = "{Anti de Sitter space and holography}",
    eprint = "hep-th/9802150",
    archivePrefix = "arXiv",
    reportNumber = "IASSNS-HEP-98-15",
    doi = "10.4310/ATMP.1998.v2.n2.a2",
    journal = "Adv. Theor. Math. Phys.",
    volume = "2",
    pages = "253--291",
    year = "1998"
}

@article{Witten:2021unn,
    author = "Witten, Edward",
    title = "{Gravity and the crossed product}",
    eprint = "2112.12828",
    archivePrefix = "arXiv",
    primaryClass = "hep-th",
    doi = "10.1007/JHEP10(2022)008",
    journal = "JHEP",
    volume = "10",
    pages = "008",
    year = "2022"
}

@inbook{Witten:2021jzq,
    author = "Witten, Edward",
    title = "{Why does quantum field theory in curved spacetime make sense? And what happens to the algebra of observables in the thermodynamic limit?}",
    eprint = "2112.11614",
    archivePrefix = "arXiv",
    primaryClass = "hep-th",
    doi = "10.1007/978-3-031-17523-7_11",
    year = "2022"
}

@article{Witten:2018zxz,
    author = "Witten, Edward",
    title = "{APS Medal for Exceptional Achievement in Research: Invited article on entanglement properties of quantum field theory}",
    eprint = "1803.04993",
    archivePrefix = "arXiv",
    primaryClass = "hep-th",
    doi = "10.1103/RevModPhys.90.045003",
    journal = "Rev. Mod. Phys.",
    volume = "90",
    number = "4",
    pages = "045003",
    year = "2018"
}

@article{Witten:2023xze,
    author = "Witten, Edward",
    title = "{A background-independent algebra in quantum gravity}",
    eprint = "2308.03663",
    archivePrefix = "arXiv",
    primaryClass = "hep-th",
    doi = "10.1007/JHEP03(2024)077",
    journal = "JHEP",
    volume = "03",
    pages = "077",
    year = "2024"
}

\end{document}